\documentclass[pre,aps,showpacs,amsmath,amssymb]{revtex4-1}
\usepackage{graphicx}
\usepackage{bm}
\usepackage{soul}
\usepackage{color}

\begin{document}

\title{Pattern formation of the three-layer Saffman-Taylor problem in a radial Hele-Shaw cell}
\author{M. Zhao$^{1}$}
\email[]{mzhao9@uci.edu}
\author{Pedro H. A. Anjos$^{2}$}
\email[]{pamorimanjos@iit.edu}
\author{J. Lowengrub$^{1}$}
\email[]{lowengrb@math.uci.edu}
\author{Shuwang Li$^{2}$}
\email[]{sli@math.iit.edu}
\affiliation{$^1$ Department of Mathematics, University of California at Irvine, Irvine, California 92697, USA\\
$^{2}$ Department of Applied Mathematics, Illinois Institute of Technology,
Chicago, Illinois  60616, USA}


\begin{abstract}

The three-layer Saffman-Taylor problem introduces two coupled moving interfaces separating the three fluids. A very recent weakly nonlinear analysis of this problem in a radial Hele-Shaw cell setup has shown that the morphologies of the emerging fingering patterns strongly depend on the initial thickness of the intermediate layer connecting the two interfaces. Here we go beyond the weakly nonlinear regime and explore full nonlinear interfacial dynamics using a spectrally accurate boundary integral method. We quantify the nonlinear instability of both interfaces as the relevant physical parameters (e.g., viscosities and surface tensions) are varied and show that our nonlinear computations are in good agreement with the experimental observations and the weakly nonlinear analysis. Nonlinear simulations reveal that due to the existence of a second interface, the classical highly branched morphologies are replaced by less unstable structures in which finger tip-splitting and finger competition phenomena are evidently restrained as the initial annulus' thickness is reduced. In addition, these novels patterns develop fingers with a series of low-amplitude bumps at their tips, associated with the enhanced growth of high-frequency modes promoted by the increasing coupling strength of interfaces. 

\end{abstract}
\maketitle

\section{Introduction}
\label{intro}

When it concerns nonlinear phenomenology, the formation and evolution of patterned structures are one of the most exciting areas. Of particular interest for physicists, engineers, and mathematicians, the hydrodynamic pattern formation of a growing interface between two fluids has captured a lot of attention over the past several decades. One of the most investigated pattern-formation systems of this type is the Saffman-Taylor problem~\cite{PG,homsy1987viscous,mccloud1995experimental,casademunt2004viscous}, which takes place when a less viscous fluid displaces a more viscous one in the confined geometry of two narrowly spaced parallel plates of Hele-Shaw cell arrangement. As the more viscous fluid is displaced, the initially flat interface between these two immiscible fluids becomes unstable and deforms, exhibiting the formation of viscous fingers.  

A very popular and widely studied Hele-Shaw cell arrangement is the radial geometry~\cite{bataille1968stabilite,Wilson,Lp}, where the less viscous fluid is injected through a central inlet and drives the more viscous fluid radially outward. As the initially slightly perturbed circular fluid-fluid interface expands, it deforms, and fingerlike protuberances form. The initial growth of the interfacial perturbations agrees with the predictions of linear stability theory~\cite{bataille1968stabilite,Wilson,Lp}. Later, the unstable modes of the flow grow and become coupled, requiring a weakly nonlinear approach to describe this regime of evolution~\cite{JM}. After that, the system evolves to a complicated late stage, characterized by the formation of convoluted fingering structures, in which nonlinear effects dominate~\cite{homsy1987viscous,mccloud1995experimental,casademunt2004viscous}. In this advance-time regime, one observes the development of visually striking, fanlike, branched patterns, in which viscous fingers compete and continue to evolve through repeated tip splitting. Due to their highly nonlinear nature, these fully nonlinear structures cannot be described theoretically by linear and weakly nonlinear approaches, and one needs to resort to sophisticated numerical techniques~\cite{ShuwangPRL,Zhao17,ShuwangJCP}.

An interesting modification of this two-fluid radial displacement consists of adding a third layer of fluid, in such a way that the system now is composed of two interfaces separating the three layers of fluids. Despite the simple modification, the inclusion of a third fluid and, consequently, a second interface, turns this system even more challenging to describe by theoretical tools since the two interfaces are already coupled at the linear level~\cite{woods,pedronew,cardoso1995formation}. Perhaps that is the reason for this three-layer radial flow has remained little explored over the years. Only a very few groups have performed studies on three-layer radial flows. Cardoso and
Woods~\cite{cardoso1995formation} carried out theoretical and experimental investigations of an immiscible three-layer radial displacement. By employing a linear stability analysis, they found a new stabilizing effect which results from the thinning of the intermediate
fluid annulus as the interfaces approach one another. This stabilizing effect, with no analogous in the single-interface problem (classical two-layer radial Hele-Shaw flow), tends to stabilize any particular perturbation to the outer interface. Beeson-Jones and Woods~\cite{woods} have studied this problem but focusing on finding the optimal value of the viscosity of the intermediate fluid in order to inject fluid at the fastest rate possible while maintaining a stable flow. Gin and Daripa have also contributed to the development of this theme~\cite{Daripa}. In Ref.~\cite{Daripa}, the authors considered a multilayer radial Hele-Shaw flow and performed a linear stability analysis of an arbitrary number of
fluid layers to design more stable injection schemes. 

More recently, Anjos and Li~\cite{pedronew} extended the perturbation theory beyond linear stability~\cite{cardoso1995formation,woods,Daripa,Daripa08} to a second-order mode-coupling theory for investigating the influence of the initial annulus' thickness $d$ on the shape of the emerging weakly nonlinear fingering patterns. Under the circumstances where the inner interface is unstable and the outer one is stable, their theoretical results indicate that as $d$ decreases, the coupling between the interfaces becomes stronger and the nearly matched final shapes exhibit the formation of wide fingers with bifurcated tips. However, if $d$ is reduced further, they observed an unexpected change in the morphology of the patterns, where the conventional finger splitting morphologies are replaced by polygonal-like structures with narrow fingers.  Although the weakly nonlinear analysis performed in Ref.~\cite{pedronew} provided insights into the morphology of the interfaces at the weakly nonlinear regime of evolution, there is still interest in unveiling the fully nonlinear features that are presented only in the advance-time regime. To do that, one necessarily needs to utilize an efficient numerical tool capable of accurately simulating the flow of the coupled dual-interface system. 

In this paper, using a boundary integral scheme we simulate the coupled dynamics of the two interfaces focusing on the fully nonlinear stage of the flow. The idea is to reformulate the differential equations into two coupled Fredholm integral equations of the second kind~\cite{LapMCD}. The boundary integral formulation reduces the dimensionality of the problem by one, i.e., the originally two-dimensional domain problem is reduced to a one-dimensional curve evolution problem. The integrals are discretized with spectral accurate quadratures~\cite{sidi1988quadrature,shelley1992study} and solved efficiently via the fast multipole method~\cite{FAST}. To remove the severe stiffness constraint from the surface tension in the Young-Laplace pressure boundary condition~\cite{PG,homsy1987viscous,mccloud1995experimental,casademunt2004viscous}, we implement the small scale decomposition idea~\cite{HLS,Zhao2015} and develop a semi-implicit time-stepping method. Our numerical algorithm is second-order accurate in time and therefore this is the first time that the coupled dual-interface system is solved accurately and efficiently.

We first compare the temporal evolution of the interfacial amplitudes given by our numerical scheme to the evolution as predicted by a second-order mode-coupling theory. Our results indicate that the fully nonlinear approach is indeed necessary to get an accurate picture of the morphological elements that arise at the later time regime of the three-layer radial Hele-Shaw flow. Then, we test the consistency of our nonlinear patterns with the experimental patterns reported in Ref.~\cite{cardoso1995formation}, showing that our nonlinear simulations are in good agreement to the experimental observations. In particular, our nonlinear computations indeed capture a unique unstable behavior--formation of drops on the outer interface while the inner one remains circular, as shown in Ref.~\cite{cardoso1995formation}.

Finally, we examine the influence of the initial annulus' thickness, which quantifies the strength of coupling between the two interfaces, on the final morphologies of the fingering patterns. It is found that the usual highly branched morphologies are replaced by less unstable structures in which finger-tip splitting and finger competition phenomena are evidently restrained as the initial annulus' thickness is reduced. In addition, these less unstable patterns develop fingers with a series of low-amplitude bumps at their tips, associated with the enhanced growth of high-frequency modes promoted by the increasing coupling strength of interfaces. We also quantify the nonlinear instability of both interfaces as the relevant physical parameters are varied, i.e., fluids' viscosities and surface tensions.  Our computations indicate that the outer interface gets increasingly more unstable as the viscosity of the innermost fluid is decreased while keeping all the other viscosities unchanged. The opposite scenario, in which the viscosity of the outermost fluid is increased while the others remain constant, points to a stabilization of the inner interface turning it less unstable.

The remainder of this paper is structured as follows. In Sec.~\ref{derivation} we present the governing equations of the dual-interface radial Hele-Shaw displacement together with a derivation of the set of coupled weakly nonlinear (WNL) equations for this system. The boundary integral scheme utilized to gain access to the interfacial shapes is also demonstrated. In Sec.~\ref{discussion} we focus on discussing our numerical results by presenting the long-time interfacial morphologies found for both inner and outer interfaces as the physical parameters are varied, with particular attention on the effects of the initial annulus' thickness on the shape of the viscous fingering patterns. A comparison between numerical and experimental patterns are also provided. Finally, in Sec.~\ref{conclude} we summarize our main results and provide some concluding remarks.

\section{Three-layer radial Hele-Shaw problem}
\label{derivation}

\subsection{Governing equations}
\label{gov}

\begin{figure}
	\centering
	\includegraphics[width=3.2 in]{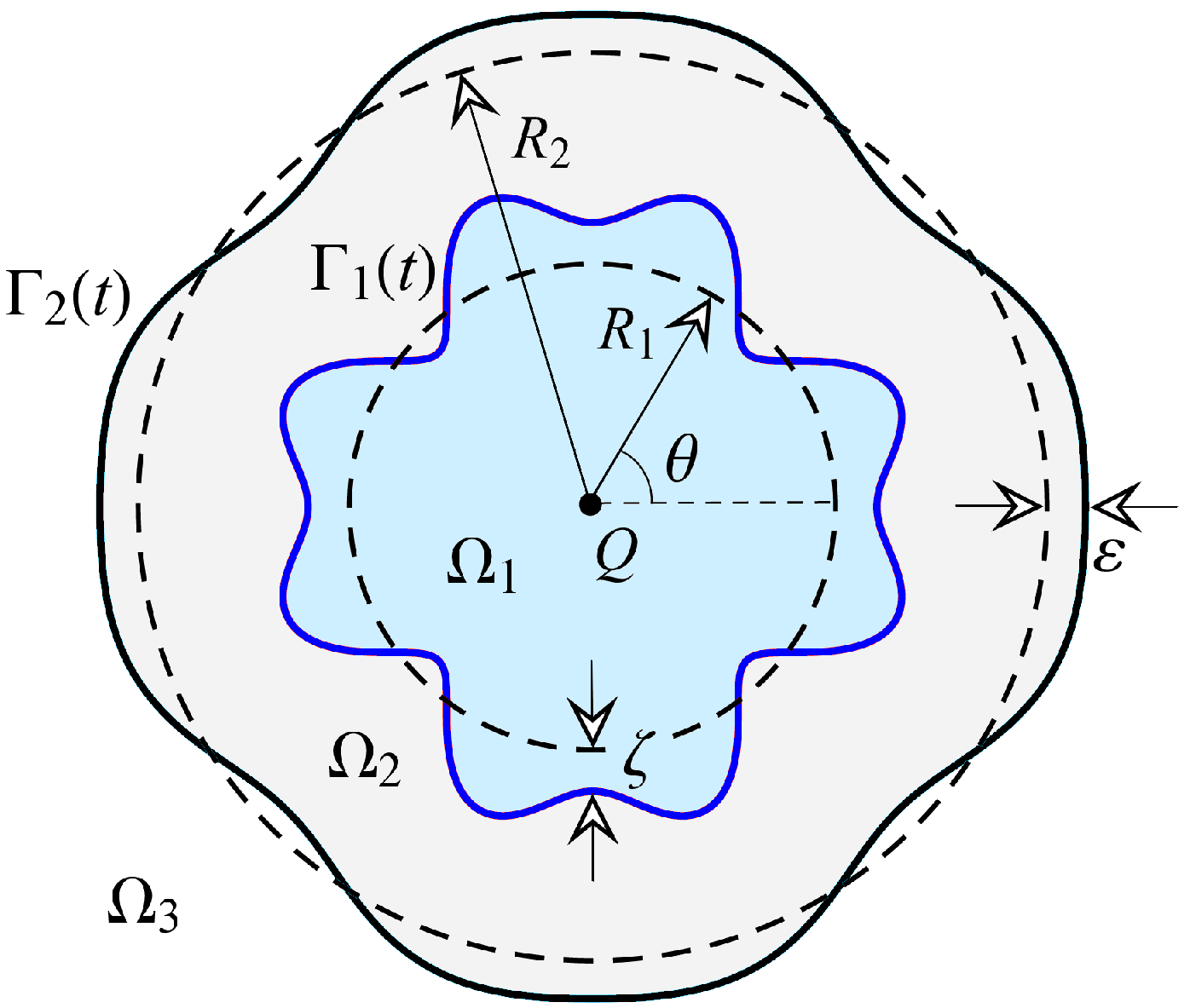}
	\caption{Schematic illustration (top view) of the injection-driven three-fluid motion in a radial Hele-Shaw cell geometry. The outermost fluid domain $\Omega_3$ and the innermost fluid domain $\Omega_1$ are separated by an annulus domain $\Omega_2$. The trailing inner interface $\Gamma_1 (t)$ is represented by the solid blue line, and the leading interface $\Gamma_2 (t)$ is depicted by the solid black line. The inner and outer dashed curves represent the time-dependent unperturbed radii of the trailing ($R_1$) and leading ($R_2$) interfaces, respectively. Fluid 1 is injected at a constant injection rate $Q$ through an inlet $\Sigma_0$ (illustrated by a small dot) located at the center of the upper plate.}
	
	\label{geom}
\end{figure}

Consider a radial Hele-Shaw cell of constant gap width $b$ containing three layers of immiscible, incompressible, Newtonian fluids, and let $\Gamma_1 (t)$ be the trailing moving interface separating fluid domain $\Omega_1$ from fluid domain $\Omega_2$. Likewise, let $\Gamma_2 (t)$ be the leading moving interface separating fluid domain $\Omega_2$ from fluid domain $\Omega_3$. See Fig.~\ref{geom} for a schematic diagram of the dual-interface system.
Here we denote the viscosity of each fluid domain as $\mu_j$ with $j=1, 2, 3$. The subscripts $1$, $2$, and $3$ refer to the inner, intermediate (annulus) and outer fluids, respectively.

For the flow between two narrowly separated plates of a Hele-Shaw cell setup, motion is governed by two equations: the gap-averaged Darcy's law~\cite{PG,homsy1987viscous,mccloud1995experimental,casademunt2004viscous,Lp}
\begin{equation}
\label{darcy}
\mathbf{u}_j=-\frac{b^2}{12\mu_j}\nabla P_j \quad \textrm{for}~ \mathbf{x}\in \Omega_{j},
\end{equation}
and a gap-averaged incompressibility condition
\begin{equation}
\label{incom}
\nabla\cdot \mathbf{u}_j=0\quad \textrm{for}~ \mathbf{x}\in \Omega_{j},
\end{equation}
where $\mathbf{u}_j$ and $P_j$ are the gap-averaged velocity and pressure of fluid $j$, respectively, and $\mathbf{x}$ denotes the position vector with origin located at the center of the cell. Due to the irrotational nature of the flow ($\nabla \times \mathbf{u}_j=0$), we can state the problem in terms of a Laplacian velocity potential $\phi_j$ ($\nabla^2 \phi_j=0$).

We also assume that fluid $1$ is injected at the origin at a constant rate
\begin{equation}
\label{flux}
{Q = \int_{\Sigma_0}\mathbf{u}_1\cdot {\mathbf{\hat{n}}}ds}
\end{equation}
and displaces fluids $2$ and $3$ radially. In Eq.~(\ref{flux}), $\Sigma_0$ is a small circle centered at origin, $s$ is the arclength, and ${\bf {\hat{n}}}$ is the outward unit normal vector. Note that the trailing and leading interfaces move radially outward with velocities $\dot R_{1}=Q/(2\pi R_{1})$ and $\dot R_{2}=Q/(2\pi R_{2})$, respectively, where $R_{1}=R_{1}(t)= \sqrt{{R_{1}^{2}(0)}+Qt/\pi}$ and $R_{2}=R_{2}(t)= \sqrt{{R_{2}^{2}(0)}+Qt/\pi}$ are the time-dependent unperturbed radii of the interfaces. Therefore,  there is a continuous thinning of the fluid annulus since $\dot R_{1} > \dot R_{2}$. As shown in previous linear~\cite{cardoso1995formation,woods,Daripa} and weakly nonlinear~\cite{pedronew} analysis of this problem, the ratio of the unperturbed radii 
\begin{equation}
\label{R}
R=R(t)=\frac{R_{1}}{R_{2}}
\end{equation} 
plays an important role in our system as it measures the coupling strength between the trailing and leading interfaces, and the initial annulus' thickness is written as $d=R_{2}(0) - R_{1}(0)$.

To include the contribution coming from surface tension, we consider the Young-Laplace pressure boundary condition~\cite{PG,homsy1987viscous,mccloud1995experimental,casademunt2004viscous}, which expresses the pressure jump across each fluid-fluid interface
\begin{equation}
\label{pre1}
P_1-P_2=\sigma_{12}\kappa_{12} \quad \textrm{for}~ \mathbf{x}\in\Gamma_1,
\end{equation}
\begin{equation}
\label{pre2}
P_2-P_3=\sigma_{23}\kappa_{23} \quad \textrm{for}~ \mathbf{x}\in\Gamma_2,
\end{equation}
where $\sigma_{12}$ ($\sigma_{23}$) is the surface tension between fluids 1 and 2 (2 and 3), and $\kappa_{12}$ ($\kappa_{23}$) denotes the interfacial curvature on the plane of the cell for $\Gamma_1$ ($\Gamma_2$).

The problem we study is specified by the pressure jump boundary condition~(\ref{pre1})-(\ref{pre2}), plus the kinematic boundary condition~\cite{homsy1987viscous,mccloud1995experimental,casademunt2004viscous} which states that the normal components of each fluid's velocity are continuous at the interfaces
\begin{equation}
\label{kine1}
\mathbf{u}_1\cdot\mathbf{\hat{n}}=\mathbf{u}_2\cdot\mathbf{\hat{n}} \quad \textrm{for}~ \mathbf{x}\in\Gamma_1, 
\end{equation}
\begin{equation}
\label{kine2}
\mathbf{u}_2\cdot\mathbf{\hat{n}}=\mathbf{u}_3\cdot\mathbf{\hat{n}} \quad \textrm{for}~ \mathbf{x}\in\Gamma_2,
\end{equation}

Now, rescaling lengths and time by $L_0=R_1(0)$ and $T_0=2 \pi R_1^2 (0)/Q$, respectively, and retaining the same notation, we have the following nondimensionalized equations:
\begin{eqnarray}
\mathbf{u}_j=-\frac{1}{\beta_{2j}}\nabla P_j \quad \textrm{for}~ \mathbf{x}\in \Omega_{j},\label{ndarcy} \\
\nabla\cdot \mathbf{u}_j=0\quad \textrm{for}~ \mathbf{x}\in \Omega_{j},\label{nincom} \\
P_1-P_2=\frac{1}{\rm Ca}\kappa_{12} \quad \textrm{for}~ \mathbf{x}\in\Gamma_1,\label{npj1} \\
P_2-P_3=\frac{\alpha}{\rm Ca}\kappa_{23} \quad \textrm{for}~ \mathbf{x}\in\Gamma_2,\label{npj2}\\
\mathbf{u}_1\cdot\mathbf{\hat{n}}=\mathbf{u}_2\cdot\mathbf{\hat{n}} \quad \textrm{for}~ \mathbf{x}\in\Gamma_1,\label{nvc1} \\
\mathbf{u}_2\cdot\mathbf{\hat{n}}=\mathbf{u}_3\cdot\mathbf{\hat{n}} \quad \textrm{for}~ \mathbf{x}\in\Gamma_2,\label{nvc2}\\
{\int_{\Sigma_0}\mathbf{u}_1\cdot {\mathbf{\hat{n}}}ds}=2\pi,\label{nflux}
\end{eqnarray}
where 
\begin{equation}
\label{cap}
{\rm Ca}=\frac{12 \mu_2 Q R_1(0)}{2 \pi \sigma_{12} b^2}
\end{equation}
is the capillary number that provides a relative measure of viscous to surface tension forces, $\beta_{2j}=\mu_j / \mu_2$ is the viscosity ratio of fluids $j$ and 2, and $\alpha = \sigma_{23} / \sigma_{12}$ is the ratio of the surface tensions.

In the nondimensionalized system, the initial annulus' thickness is written as
\begin{equation}
\label{d}
d=\frac{1}{R_{0}}-1,
\end{equation} 
where $R_{0}=R(t=0)$. In this way, we can conveniently study the effects of the annulus' thickness by just varying the values of the initial ratio of the unperturbed radii $R_{0}$.

\subsection{Weakly nonlinear (WNL) equations}
\label{WNL}

This section is devoted to a brief derivation of a set of second-order mode-coupling differential equations, which allows one to describe the time evolution of the interfacial perturbation amplitudes
for both inner and outer interfaces. This set of nonlinear differential equations, originally obtained in Ref.~\cite{pedronew}, permits the analytical investigation of the role played by the initial annulus' thickness in regulating the overall morphologies of the viscous fingering patterns up to the weakly nonlinear regime. Nevertheless, in the current study, we focus on the analysis of dynamical evolution of the dual-interface system in the fully nonlinear regime and therefore the perturbative weakly nonlinear equations will only be utilized in the comparison with the fully nonlinear numerical solution given by our boundary integral formulation (see Sec.~\ref{boundary}). We direct the interested readers to Ref.~\cite{pedronew} for a detailed discussion about the mode-coupling strategy, its description and proposed interpretation, as well as its usage on describing the weakly nonlinear regime of the injection-driven three-layer radial Hele-Shaw
flows.

During the injection process, the initially slightly perturbed, circular interfaces can become unstable, and deform, due to the interplay of viscous and capillary forces. Therefore, we express the perturbed trailing interface as ${\cal R}_{1} ={\cal R}_{1}(\theta,t)= R_{1}(t) + \zeta(\theta,t)$, where $\theta$ denotes the azimuthal angle in the $r-\theta$ plane. The radial coordinate $r$ denotes the distance to the injection source point,
which is chosen as the origin of the polar coordinate system. The net interface disturbance is represented as a Fourier series
\begin{equation}
\label{fourier1}
\zeta(\theta,t)=\sum_{n=-\infty}^{+\infty} \zeta_{n}(t)~e^{in\theta}, 
\end{equation}
where $\zeta_{n}(t)$ denotes the complex Fourier amplitudes, with integer wave numbers $n$. Likewise, we represent the perturbed leading interface as ${\cal R}_{2} ={\cal R}_{2}(\theta,t)= R_{2}(t) + \varepsilon(\theta,t)$, and
\begin{equation}
\label{fourier2}
\varepsilon(\theta,t)=\sum_{n=-\infty}^{+\infty} \varepsilon_{n}(t)~e^{in\theta}. 
\end{equation}
Our perturbative approach keeps terms up to the second order in $\zeta$ and $\varepsilon$.

At this point, we have all ingredients needed to derive the set of mode-coupling differential equations for the perturbation amplitudes $\zeta_{n}(t)$ and $\varepsilon_{n}(t)$. 
Following the steps performed in the weakly nonlinear study of Ref.~\cite{pedronew}, we perform Fourier expansions for the velocity potentials $\phi_j$, and use the kinematic boundary condition [Eqs.~(\ref{nvc1}) and (\ref{nvc2})] 
to express the Fourier coefficients of $\phi_{j}$ in terms of $\zeta_{n}$ and $\varepsilon_{n}$. Substituting these relations, and the pressure jump condition 
[Eqs.~(\ref{npj1}) and (\ref{npj2})] into the Darcy's law [Eq.~(\ref{ndarcy})] and keeping terms up to second-order in $\zeta$ and $\varepsilon$, we obtain the set of dimensionless coupled equations of motion for both perturbation amplitudes $\zeta_{n}$ and $\varepsilon_{n}$ (for $n \ne 0$)

\begin{eqnarray}
\label{result1}
\dot{\zeta}_{n} &=& \underbrace{f_{1}~\Bigg[\frac{|n|-f_{1}^{-1}}{R_{1}^2} - \left(\frac{1}{1-\beta_{21}}\right) \frac{|n|(n^2-1)}{{\rm Ca}R_{1}^{3}}\Bigg]{\zeta}_{n} + f_{2}~\Bigg[\frac{|n|}{R_{2}^2} - \left(\frac{1}{\beta_{23}-1}\right) \frac{|n|(n^2-1)}{{\rm Ca}R_{2}^{3}}\Bigg]{\varepsilon}_{n}}_{\text{linear}} \nonumber\\
&+& \bigg \{ \bigg. f_{1}\sum_{n' \neq 0} \Big [F(n, n')\zeta_{n'}^{{\rm lin}} \zeta_{n - n'}^{{\rm lin}} + G(n, n')\dot{\zeta}_{n'}^{{\rm lin}} \zeta_{n - n'}^{{\rm lin}} \Big ] + f_{2}\sum_{n' \neq 0} \Big [H(n, n')\varepsilon_{n'}^{{\rm lin}} \varepsilon_{n - n'}^{{\rm lin}} + I(n, n')\dot{\varepsilon}_{n'}^{{\rm lin}} \varepsilon_{n - n'}^{{\rm lin}} \Big ] \nonumber\\
&+& \underbrace{f_{2}\sum_{n' \neq 0} \Big [ \Big. J(n, n')\varepsilon_{n'}^{{\rm lin}} \zeta_{n - n'}^{{\rm lin}} + K(n, n')\dot{\varepsilon}_{n'}^{{\rm lin}} \zeta_{n - n'}^{{\rm lin}} + L(n, n')\zeta_{n'}^{{\rm lin}} \varepsilon_{n - n'}^{{\rm lin}} + M(n, n')\dot{\zeta}_{n'}^{{\rm lin}} \varepsilon_{n - n'}^{{\rm lin}} \Big. \Big ]\bigg. \bigg \},~~~~~~~~~~}_{\text{WNL}}
\end{eqnarray}
\begin{eqnarray}
\label{result2}
\dot{\varepsilon}_{n} &=& \underbrace{f_{3}~\Bigg[\frac{|n|}{R_{1}^2} - \left(\frac{1}{1 - \beta_{21}}\right) \frac{|n|(n^2-1)}{{\rm Ca}R_{1}^{3}}\Bigg]{\zeta}_{n} +f_{4}~\Bigg[\frac{|n|-f_{4}^{-1}}{R_{2}^2} - \left(\frac{1}{\beta_{23}-1}\right) \frac{|n|(n^2-1)}{{\rm Ca}R_{2}^{3}}\Bigg]{\varepsilon}_{n}}_{\text{linear}}  \nonumber\\
&+& \bigg \{ \bigg. f_{3}\sum_{n' \neq 0} \Big [{\cal F}(n, n')\zeta_{n'}^{{\rm lin}} \zeta_{n - n'}^{{\rm lin}} + {\cal G}(n, n')\dot{\zeta}_{n'}^{{\rm lin}} \zeta_{n - n'}^{{\rm lin}} \Big ] + f_{4}\sum_{n' \neq 0} \Big [{\cal H}(n, n')\varepsilon_{n'}^{{\rm lin}} \varepsilon_{n - n'}^{{\rm lin}} + {\cal I}(n, n')\dot{\varepsilon}_{n'}^{{\rm lin}} \varepsilon_{n - n'}^{{\rm lin}} \Big ] \nonumber\\
&+& 
\underbrace{f_{3}\sum_{n' \neq 0} \Big [ \Big. {\cal J}(n, n')\varepsilon_{n'}^{{\rm lin}} \zeta_{n - n'}^{{\rm lin}} + {\cal K}(n, n')\dot{\varepsilon}_{n'}^{{\rm lin}} \zeta_{n - n'}^{{\rm lin}} + {\cal L}(n, n')\zeta_{n'}^{{\rm lin}} \varepsilon_{n - n'}^{{\rm lin}} + {\cal M}(n, n')\dot{\zeta}_{n'}^{{\rm lin}} \varepsilon_{n - n'}^{{\rm lin}} \Big. \Big ]\bigg. \bigg \}.~~~~~~~~~~}_{\text{WNL}}
\end{eqnarray}
In Eqs.~(\ref{result1}) and~(\ref{result2}),
\begin{eqnarray}
f_{1} &=& \frac{A_{12} (1-A_{23}R^{2|n|})}{1+A_{12}A_{23}R^{2|n|}} ,\  ~~~~~~~~~~f_{2} = \frac{A_{23} (1+A_{12})R^{(|n|-1)}}{1+A_{12}A_{23}R^{2|n|}},\nonumber\\
f_{3} &=& \frac{A_{12} (1-A_{23})R^{(|n|+1)}}{1+A_{12}A_{23}R^{2|n|}},\  ~~~~~~~~f_{4} = \frac{A_{23} (1+A_{12}R^{2|n|})}{1+A_{12}A_{23}R^{2|n|}},\nonumber
\label{parameter}
\end{eqnarray}
where $A_{12}=(1-\beta_{21})/(1+\beta_{21})$ [$A_{23}=(\beta_{23}-1)/(\beta_{23}+1)$] is the viscosity contrast of fluids 1 and 2 (2 and 3) written in terms of the viscosity ratios. Moreover, these equations are obtained considering $\alpha=1$. The expressions for the second-order mode-coupling terms are given in the Appendix A [see Eqs.~(\ref{F1})-(\ref{g4})]. 

Note that in the limit of a thick annulus ($R \rightarrow 0$), the terms multiplied by the functions $f_2$ [in Eq.~(\ref{result1})] and $f_3$ [in Eq.~(\ref{result2})] become weak, leaving two decoupled single interfaces. In this case, and after appropriate reintroduction of dimensions, it can be shown that each one of these equations reduces to the considerably simpler expression obtained in Ref.~\cite{JM} for the corresponding two-fluid (single-interface) version of the problem. On the other hand, in the limit of a thin annulus ($R \rightarrow 1$, $R_1 \rightarrow R_2$, and $\zeta_{n} \rightarrow \varepsilon_{n}$) both equations reduce to an expression describing the radial displacement of fluid 3 directly by fluid 1, but with interfacial tension equal to the sum of the interfacial tensions of the two interfaces ($\sigma_{12}+\sigma_{23}=2\sigma$). It should be pointed out that, in order to make the weakly nonlinear equations consistent with our boundary integral scheme, Eqs.~(\ref{result1}) and~(\ref{result2}) are made dimensionless by rescaling lengths and time by $L_0=R_1(0)$ and $T_0=2 \pi R_1^2 (0)/Q$, respectively. This rescaling choice utilized to nondimensionalize our equations is a bit different from the one performed in Ref.~\cite{pedronew}.

\subsection{Boundary integral formulation}
\label{boundary}

Considering the Darcy's law Eq.~(\ref{ndarcy}), we can take the velocity potential as $\displaystyle \phi_j=-\frac{P_j}{\beta_{2j}}$ to be a dimensionless potential function. From Eqs.~(\ref{ndarcy})-(\ref{nvc2}), the potential functions are harmonic and have continuous normal derivatives across both interfaces. Thus,
the velocity potential $\phi$ satisfies a double layer potential,
\begin{equation}
\phi(\mathbf{x})=\frac{1}{2\pi}\int_{\Gamma_1}\gamma_1\frac{\partial \ln |\mathbf{x}-\mathbf{x}'|}{\partial \mathbf{\hat{n}(x')}}d s(\mathbf{x'})+\frac{1}{2\pi}\int_{\Gamma_2}\gamma_2\frac{\partial \ln |\mathbf{x}-\mathbf{x}'|}{\partial \mathbf{\hat{n}(x')}}d s(\mathbf{x'})+\ln |\mathbf{x}|,\label{eqphi}
\end{equation}
where $\gamma_1$ is the dipole density on the inner interface $\Gamma_1$ and $\gamma_2$ is the dipole density on the outer interface $\Gamma_2$. 

On the boundary $\Gamma_1$, the pressure jump Eq.~(\ref{npj1}) gives us
\begin{eqnarray}
\frac{1}{2}(\beta_{21}+1)\gamma_1+\frac{1}{2\pi}(\beta_{21}-1) \bigg ( \bigg. \int_{\Gamma_1}\gamma_1\frac{\partial \ln |\mathbf{x}-\mathbf{x}'|}{\partial \mathbf{\hat{n}(x')}}d s(\mathbf{x'})+\int_{\Gamma_2}\gamma_2\frac{\partial \ln |\mathbf{x}-\mathbf{x}'|}{\partial \mathbf{\hat{n}(x')}}d s(\mathbf{x'})+ 2\pi\ln |\mathbf{x}|\bigg ) \bigg.
=-\frac{1}{\rm Ca}\kappa_{12}.\label{eqmu1}
\end{eqnarray}
Similarly, on the boundary $\Gamma_2$, we use the pressure jump Eq.~(\ref{npj2}) to obtain
\begin{eqnarray}
\frac{1}{2}(\beta_{23}+1)\gamma_2+\frac{1}{2\pi}(1-\beta_{23})\bigg ( \bigg.\int_{\Gamma_1}\gamma_1\frac{\partial \ln |\mathbf{x}-\mathbf{x}'|}{\partial \mathbf{\hat{n}(x')}}d s(\mathbf{x'})+\int_{\Gamma_2}\gamma_2\frac{\partial \ln |\mathbf{x}-\mathbf{x}'|}{\partial \mathbf{\hat{n}(x')}}d s(\mathbf{x'})+ 2\pi\ln |\mathbf{x}|\bigg ) \bigg.
=-\frac{\alpha}{\rm Ca}\kappa_{23}.\label{eqmu2}
\end{eqnarray}

The Eqs.~(\ref{eqmu1}) and (\ref{eqmu2}) are well defined $2^{\rm nd}$ kind Fredholm integral equations, which can be solved via GMRES~\cite{GMRES}. Once $\gamma_1$ and $\gamma_2$ are determined, we are able to compute the normal velocities via Dirichlet-Neumann mapping~\cite{LapMCD}
\begin{eqnarray}
V_{\Gamma_1}=\frac{1}{2\pi}\int_{\Gamma_1}\gamma_{1,s'}\frac{(\mathbf{x}-\mathbf{x}')^\perp\cdot\mathbf{\hat{n}(x)}}{|\mathbf{x}-\mathbf{x}'|^2}ds'(\mathbf{x'})+\frac{1}{2\pi}\int_{\Gamma_2}\gamma_{2,s'}\frac{(\mathbf{x}-\mathbf{x}')^\perp\cdot\mathbf{\hat{n}(x)}}{|\mathbf{x}-\mathbf{x}'|^2}ds'(\mathbf{x'})+ \frac{\mathbf{x}\cdot\mathbf{\hat{n}}}{|\mathbf{x}|^2},\label{V1}\\
V_{\Gamma_2}=\frac{1}{2\pi}\int_{\Gamma_1}\gamma_{1,s'}\frac{(\mathbf{x}-\mathbf{x}')^\perp\cdot\mathbf{\hat{n}(x)}}{|\mathbf{x}-\mathbf{x}'|^2}ds'(\mathbf{x'})+\frac{1}{2\pi}\int_{\Gamma_2}\gamma_{2,s'}\frac{(\mathbf{x}-\mathbf{x}')^\perp\cdot\mathbf{\hat{n}(x)}}{|\mathbf{x}-\mathbf{x}'|^2}ds'(\mathbf{x'})+\frac{\mathbf{x}\cdot\mathbf{\hat{n}}}{|\mathbf{x}|^2},\label{V2}
\end{eqnarray}
where the subscript $s$ denotes the partial derivatives with respect to arclength $s$ and ${\textbf{x}}^{\perp}=(x_2,-x_1)$. Following Ref.~\cite{HLS}, we discretize Eqs.~(\ref{eqmu1}) and (\ref{eqmu2}) via spectrally accurate equal arclength discretization and evaluate integrals through a fast multipole method~\cite{FAST}. The discretized system of $\gamma_1$ and $\gamma_2$ is solved via an iterated method GMRES~\cite{GMRES}. We compute the normal velocity of each interface via the spectrally accurate discretization~\cite{HLS}.

Next, we evolve each interface  through
\begin{eqnarray}
\frac{d {\mathbf{x}_1}}{d {t}}\cdot \mathbf{\hat{n}_1}={V_{\Gamma_1}}&,~~~~&
\frac{d {\mathbf{x}_2}}{d {t}}\cdot \mathbf{\hat{n}_2}={V_{\Gamma_2}},
\end{eqnarray}
where calculation points ${\mathbf{x}_1} \in \Gamma_1$, ${\mathbf{x}_2} \in \Gamma_2$, and similarly for the normal vectors $\mathbf{\hat{n}_1}$ and $\mathbf{\hat{n}_2}$. This system is very stiff due to the higher-order terms introduced by the curvature and requires a severe third-order time step constraint $\Delta t\sim h^3$, where $\Delta t$ is the time step and $h$ is the spatial grid size. Following the small scale decomposition~\cite{HLS,Zhao2015}, we remove the stiffness and obtain a second-order accurate updating scheme in time. In the Appendix B we test the convergence of our scheme. Note that the small scale decomposition idea has also been successfully implemented in solving the dynamics of inextensible vesicles~\cite{Kai2014,Kai2017} and precipitate evolution in an elastic media~\cite{Amlan2014,Amlan2016}.

\section{Numerical results}
\label{discussion}

\subsection{Comparison between fully nonlinear simulations and weakly nonlinear perturbation theory}
\label{discussion2}

In this section we present a comparison between our numerical results of Sec.~\ref{boundary} and the predictions of a second-order mode-coupling theory~\cite{pedronew} presented in Sec.~\ref{WNL}. We consider the nonlinear coupling between just two Fourier cosine modes, namely, $n$ and $2n$, and utilize the weakly nonlinear Eqs.~(\ref{result1}) and (\ref{result2}) to find how the cosine amplitudes $a_{n}(t)$, $a_{2n}(t)$, $b_{n}(t)$, and $b_{2n}(t)$ evolve in time. Here $a_{n}(t)$ denotes the cosine perturbation amplitude of the fundamental mode for the inner interface and $a_{2n}(t)$ is its first harmonic. Likewise, $b_{n}(t)$ and $b_{2n}(t)$ are cosine amplitudes related to the outer interface.

\begin{figure}[t]
	\begin{center}
	\includegraphics[scale=0.37]{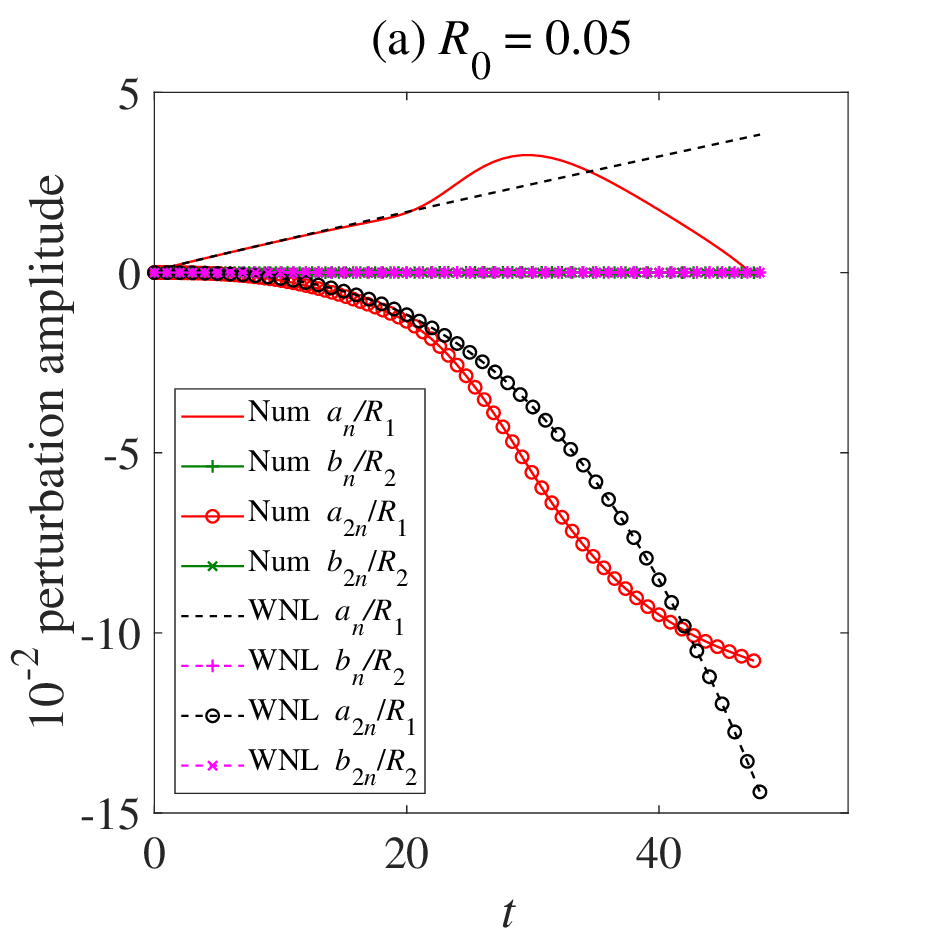}
	\includegraphics[scale=0.37]{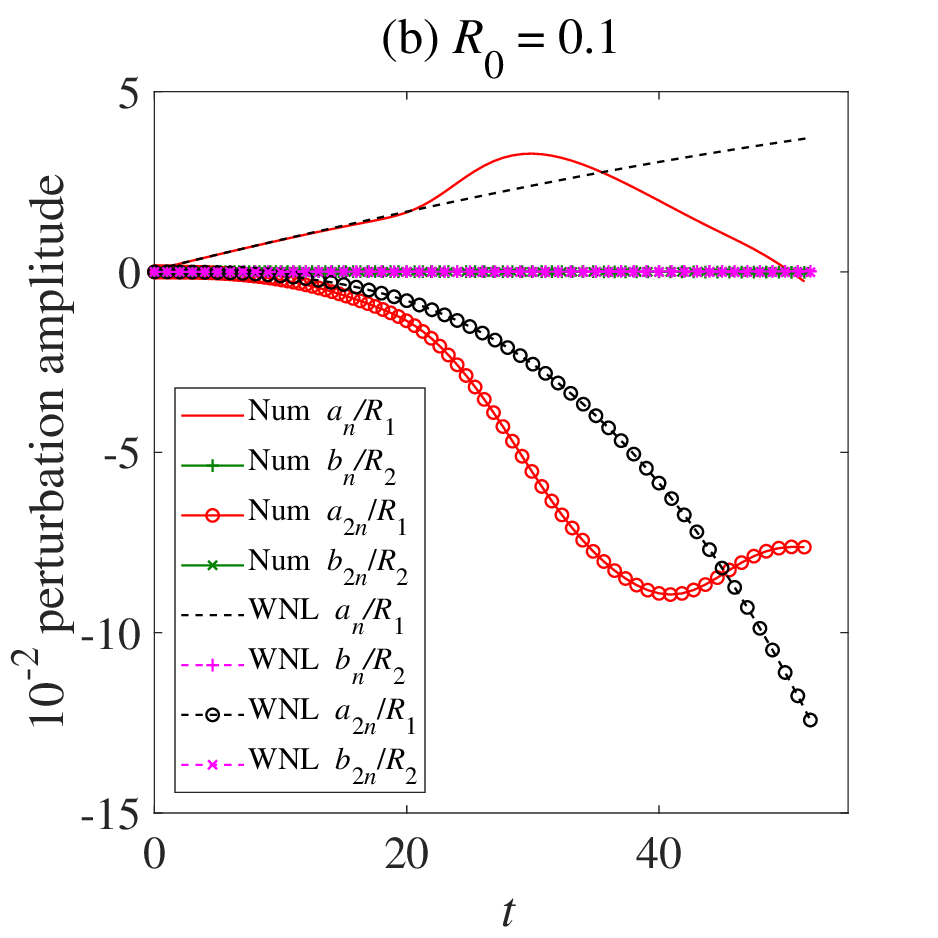}
	\includegraphics[scale=0.37]{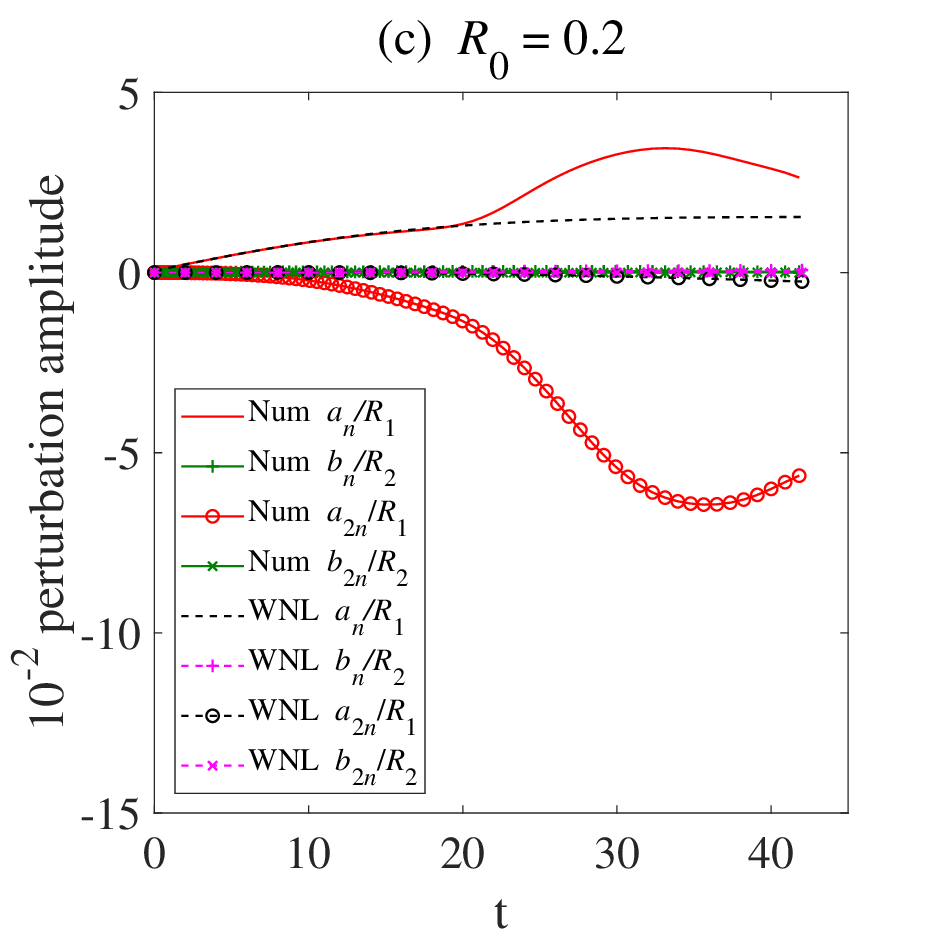}
	\end{center}

	\caption{Temporal evolution of the rescaled cosine amplitudes $a_{n}(t)/R_{1}(t)$, $a_{2n}(t)/R_{1}(t)$, $b_{n}(t)/R_{2}(t)$, and $b_{2n}(t)/R_{2}(t)$, for three increasing values of parameter $R_0$: (a) $R_{0}=0.05$, (b) $R_{0}=0.1$, and (c) $R_{0}=0.2$. The solid curves represent the time evolution given by the numerical approach of Sec.~\ref{boundary} while the dashed lines are the temporal evolution as predicted by the WNL theory of Sec.~\ref{WNL}. The initial conditions for the inner and outer interfaces are ${\cal R}_{1}(\theta,t=0)=1+5\times 10^{-4}\cos(n\theta)$ and ${\cal R}_{2}(\theta,t=0)=1/R_0+5\times 10^{-4}\cos(n\theta)$, respectively, where $n=4$. In addition, we set $\beta_{21}=0.01$, $\beta_{23}=100$, $\alpha=1$, and ${\rm Ca}=1000$. In the case of the fully nonlinear numerical amplitudes, we utilized $N=8192$ points along each interface and time step $\Delta t=1\times 10^{-3}$.}
	\label{comparison}
\end{figure}

We start our comparison by examining Fig.~\ref{comparison}, which presents the temporal evolution of the rescaled perturbation amplitudes $a_{n}(t)/R_{1}(t)$, $a_{2n}(t)/R_{1}(t)$, $b_{n}(t)/R_{2}(t)$, and $b_{2n}(t)/R_{2}(t)$, for three increasing values of parameter $R_0$: (a) $R_{0}=0.05$, (b) $R_{0}=0.1$, and (c) $R_{0}=0.2$. The solid curves represent the time evolution given by the numerical approach of Sec.~\ref{boundary} while the dashed lines are the temporal evolution as predicted by the WNL theory of Sec.~\ref{WNL}. The initial conditions for the inner and outer interfaces are ${\cal R}_{1}(\theta,t=0)=1+5\times 10^{-4}\cos(n\theta)$ and ${\cal R}_{2}(\theta,t=0)=1/R_0+5\times 10^{-4}\cos(n\theta)$, respectively, where $n=4$. In addition, we set $\beta_{21}=0.01$, $\beta_{23}=100$, $\alpha=1$, and ${\rm Ca}=1000$. In the case of the fully nonlinear numerical amplitudes, we utilized $N=8192$ points along each interface and time step $\Delta t=1\times 10^{-3}$.

The discussion is initiated by surveying Fig.~\ref{comparison}(a), in which $R_0=0.05$. It is apparent that the agreement between numerical and WNL amplitudes is excellent for the early stages of the dynamics, in which nonlinear effects are still not significant. This agreement holds up until time $t=20$ when we start to observe a difference between the fully nonlinear and WNL evolutions. At the end of the temporal evolution, we note that the fully nonlinear value of $a_{n}(t)/R_{1}(t)$ [$a_{2n}(t)/R_{1}(t)$] is smaller [larger] than the one given by the WNL evolution. 

Very similar conclusions can be drawn when $R_0=0.1$, case depicted in Fig.~\ref{comparison}(b). Now, the distance between fully nonlinear and WNL evolutions gets larger. To understand such behavior, one needs to recall that $R_0$ measures the initial coupling strength between the inner and outer interfaces. Since we are considering a larger value of $R_0$, it is expected an enhanced coupling between the interfaces followed by the growth of higher-order harmonics (such as $3n$, $4n$, etc), and therefore nonlinear effects should be more intense. The fully nonlinear numerical evolution illustrated in Fig.~\ref{comparison} takes into account the full coupling of modes in the Fourier decomposition of the perturbation. On the other hand, the WNL evolution only addresses the nonlinear coupling between the Fourier modes $n$ and $2n$, and all the other harmonics are absent. In this way, although a second-order mode-coupling theory can correctly dictate the signs of modes $n$ and $2n$, it simply cannot accurately predict the values of these perturbations amplitudes for later times of the dynamics due to the growth of higher-order harmonics.

Last, we investigate Fig.~\ref{comparison}(c) for $R_0=0.2$. Although there is a good agreement for the amplitude $a_{n}(t)/R_{1}(t)$ up until $t=20$, the same is not observed for the amplitude $a_{2n}(t)/R_{1}(t)$. While the WNL evolution predicts just a small negative growth of that amplitude, the fully nonlinear evolution dictates a strongly negative growth and the distance between these two types of evolution is even larger than the one illustrated in Fig.~\ref{comparison}(b) for $R_0=0.1$. These findings reinforce the necessity of the usage of our fully nonlinear boundary integral approach in order to get a more complete picture of the morphological elements that arise at the later time regime of the three-layer radial Hele-Shaw flow. As a last comment about Fig.~\ref{comparison}, we would like to mention that for the final time considered here, we did not observe the growth of any outer interface perturbation amplitudes [$b_{n}(t)$ and $b_{2n}(t)$] in both fully nonlinear and WNL evolutions. 

\subsection{Contrasting numerical and experimental patterns}

Before starting to analyze the role of the controlling parameters ($R_0$, $\beta_{21}$, $\beta_{23}$, $\alpha$) in the nonlinear patterns given by our numerical scheme, we would like to confirm that our numerical solutions are consistent with available experimental observations of the three-layer radial Hele-Shaw flow. Here we contrast our nonlinear simulations with an experimental situation carried out by Cardoso and Woods in Ref.~\cite{cardoso1995formation}. 

\begin{figure}[t]

\includegraphics[scale=0.27]{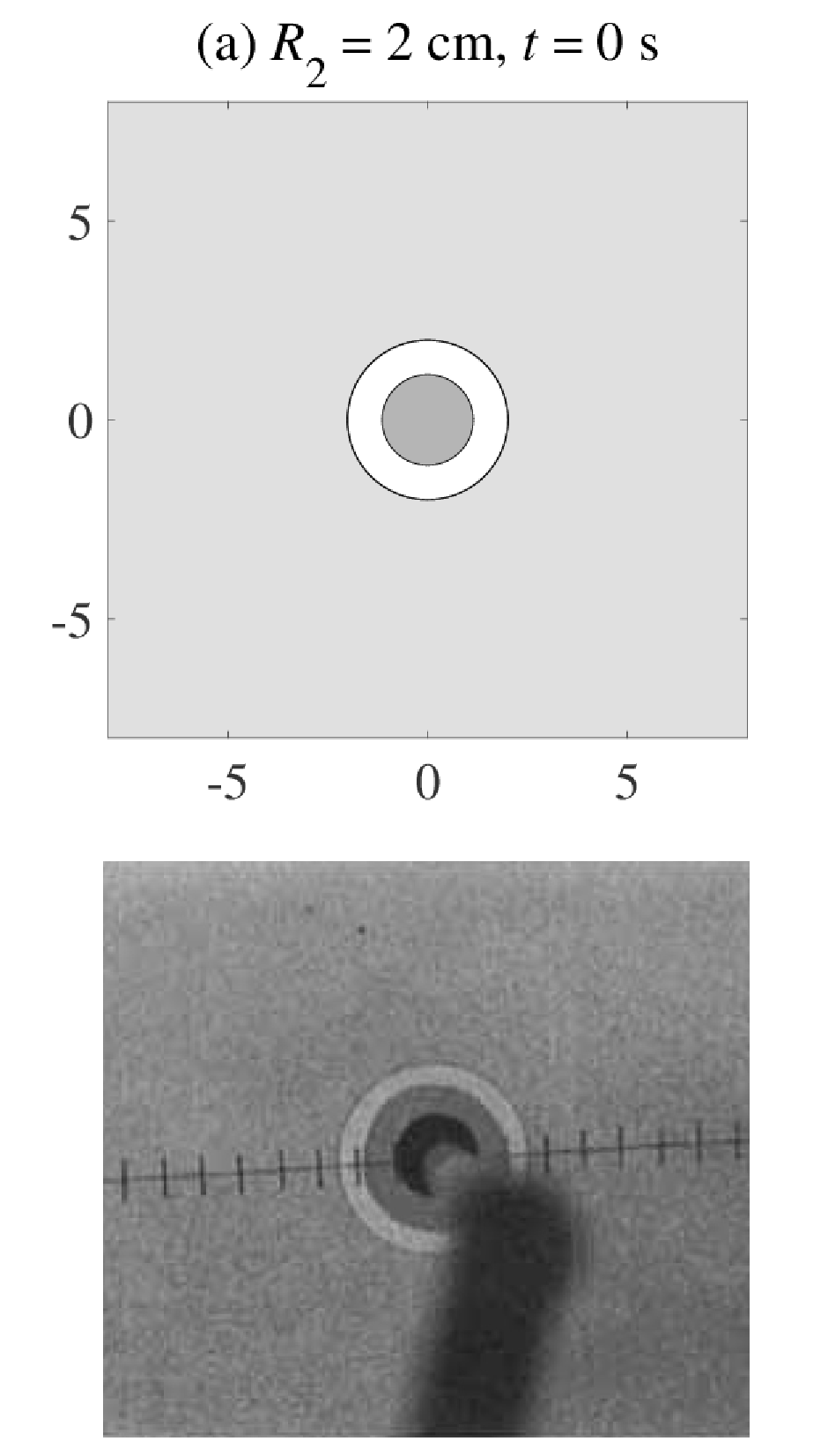}
\includegraphics[scale=0.27]{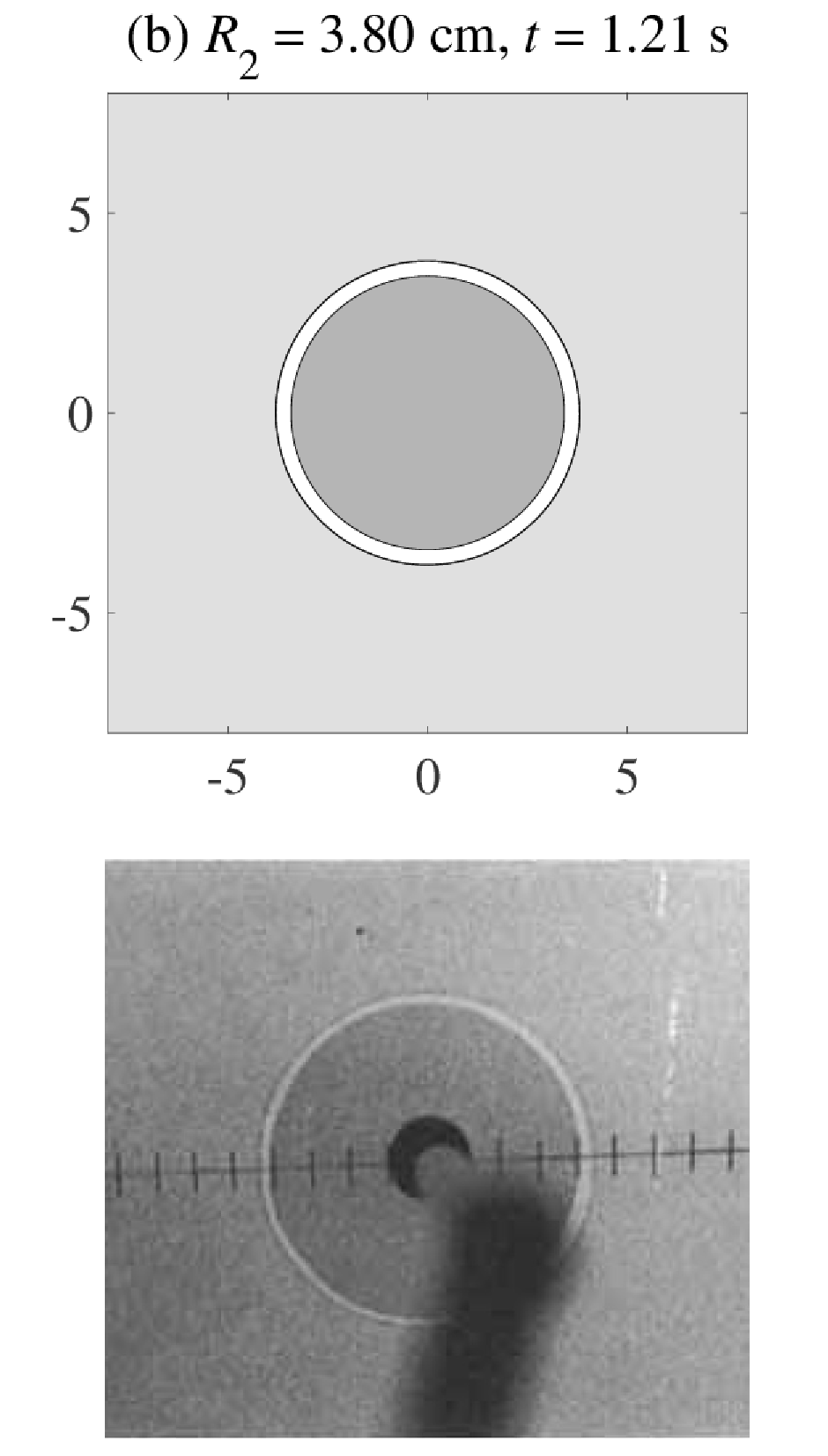}
\includegraphics[scale=0.27]{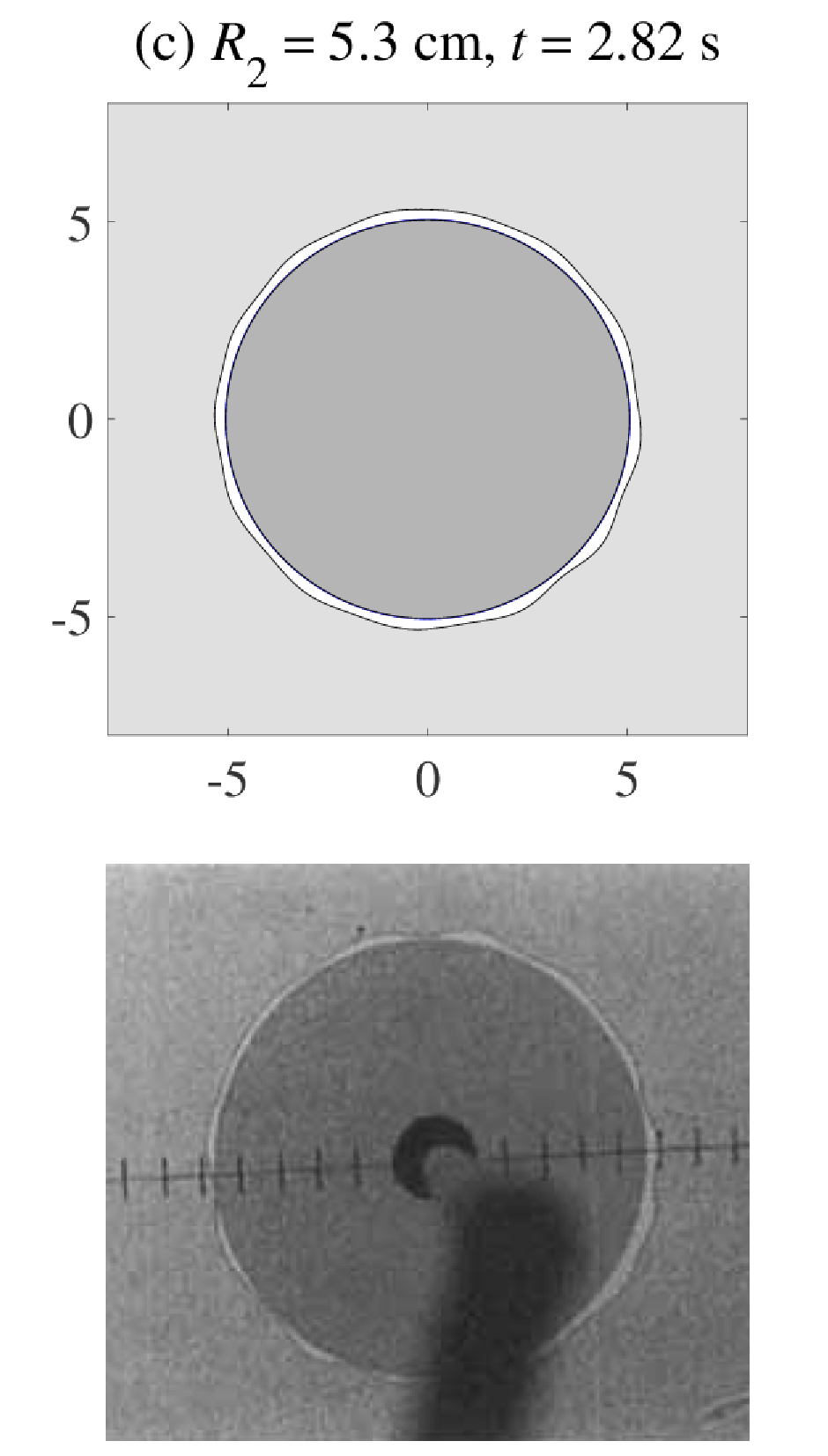}
\includegraphics[scale=0.27]{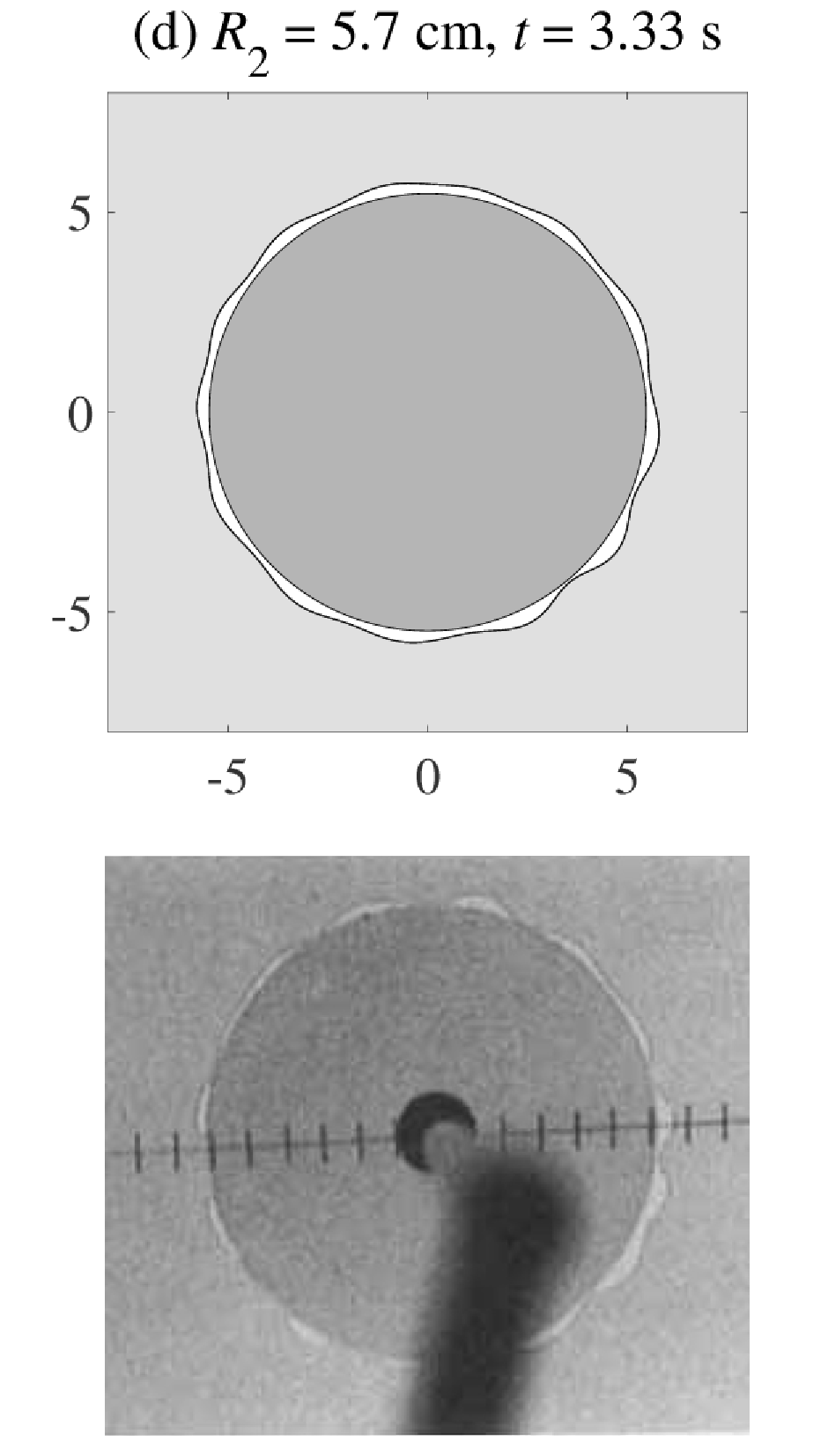}
	\caption{Comparison between experimental results (bottom panels) from Fig. 19 of Ref.~\cite{cardoso1995formation} and numerical simulations (top panels). The inner fluid 1 (natrosol solution) is injected into the annulus fluid 2 (air) at a flow rate $Q=27 \text{ cm}^2\cdot \text{s}^{-1}$, and the outer fluid 3 is cooking oil. The nonlinear interfacial morphologies and experimental patterns are shown at outer interface size: (a) $R_2=2 \text{ cm}$, (b) $R_2=3.8 \text{ cm}$, (c) $R_2=5.3\text{ cm}$, and (d) $R_2=5.7 \text{ cm}$. The scale of the experimental results is $1\text{ cm}$. Experimental results reprinted from Fig. 19 in Ref.~\cite{cardoso1995formation} with the permission of Cambridge University Press.}\label{expfig}
\end{figure}

The  case investigated in Ref.~\cite{cardoso1995formation} is depicted in Fig.~\ref{expfig}, in which natrosol solution is utilized as the first layer with mobility $M_1=5.019\times 10^{-6}\text{ cm}^3\cdot \text{s}\cdot \text{g}^{-1}$, air as the second layer with mobility $M_2=2.6\text{ cm}^3\cdot \text{s}\cdot \text{g}^{-1}$, and cooking oil as the third layer with mobility $M_3=7.24\times 10^{-4}\text{ cm}^3\cdot \text{s}\cdot \text{g}^{-1}$. The injection rate is $Q=27 \text{ cm}^2\cdot \text{s}^{-1}$ and the surface tensions are $\sigma_{12}= 66 \text{ g}\cdot \text{s}^{-2}$ and $\sigma_{23}= 32 \text{ g}\cdot \text{s}^{-2}$. The initial radius of the inner and outer interfaces are $R_1(0)=1.14\text{ cm}$ and $R_2(0)=2 \text{ cm}$, respectively. Using  length scale  $L_0=R_1(0)=1.14\text{ cm}$ and time scale $T_0=2\pi R_1^2(0)/Q=0.3024 \text{ s}$, we have nondimensional parameters $R_2(0)=1.754$, $\displaystyle\beta_{21}=\frac{M_2}{M_1}=5.22\times 10^6$, $\displaystyle \beta_{23}=\frac{M_2}{M_3}=3.6\times 10^3$, $\displaystyle {\rm Ca}=\frac{Q R_1(0)}{2 \pi \sigma_{12} M_2}=2.85\times 10^{-2}$, and $\displaystyle \alpha=\frac{\sigma_{23}}{\sigma_{12}}=0.485$. Since there is no detailed information on the initial interface morphologies given in Ref.~\cite{cardoso1995formation}, here we simply set the initial interfaces to be slightly perturbed circles 
\begin{eqnarray}
{\cal R}_{1}(\theta,0)&=&1+1\times 10^{-4}\sum\limits_{n=2}^{25}e^{-0.2n}[a_{1n}\cos(n\theta)+b_{1n}\sin(n\theta)],\\
{\cal R}_{2}(\theta,0)&=&1.754+1\times 10^{-4}\sum\limits_{n=2}^{25}e^{-0.2n}[a_{2n}\cos(n\theta)+b_{2n}\sin(n\theta)].
\end{eqnarray}

In Fig.~\ref{expfig}, the characteristic interfacial morphologies are shown at outer interface size: (a) $R_2=2 \text{ cm}$, (b) $R_2=3.8 \text{ cm}$, (c) $R_2=5.3\text{ cm}$, and (d) $R_2=5.7 \text{ cm}$ for both simulations (top) and experiments (bottom). Initially, both interfaces are slightly perturbed circles. Since the inner fluid is very viscous ($\beta_{21}\gg 1$), the inner interface is almost rigid and it remains in a circular shape as it expands. On the other hand, the outer interface (air-oil interface) develops multiple fingers as it moves outward. The outer interface emerges about 13 small fingers, the same as those in the experiment. Although our nonlinear simulation is not able to show the rupture of the annulus, as found experimentally by Cardoso and Woods~\cite{cardoso1995formation} for subsequent times, it indeed captures the onset of the airdrops formation. By comparing the top panels with the bottom ones in Fig.~\ref{expfig}, it is clear that there is a close resemblance between our numerical patterns and the experimental findings, and the agreement is excellent.

\subsection{Nonlinear fingering patterns}
\label{discussion3}

In this section, we utilize the boundary integral method of Sec.~\ref{boundary} to compute the fully nonlinear interfaces that arise during the three-layer radial Hele-Shaw displacement and to analyze the role of the controlling parameters ($R_0$, $\beta_{21}$, $\beta_{23}$, $\alpha$).

\subsubsection{Effects of $R_0$}
\label{r0}

\begin{figure}[t]
	\includegraphics[scale=0.27]{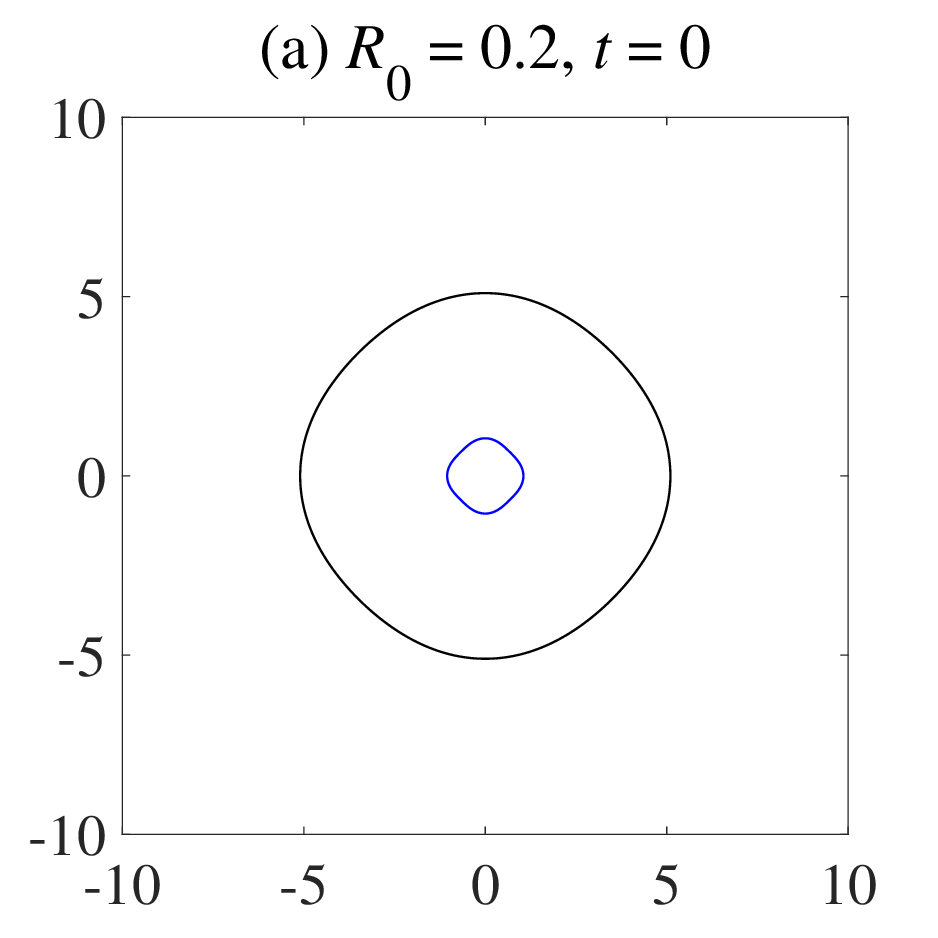}
	\includegraphics[scale=0.27]{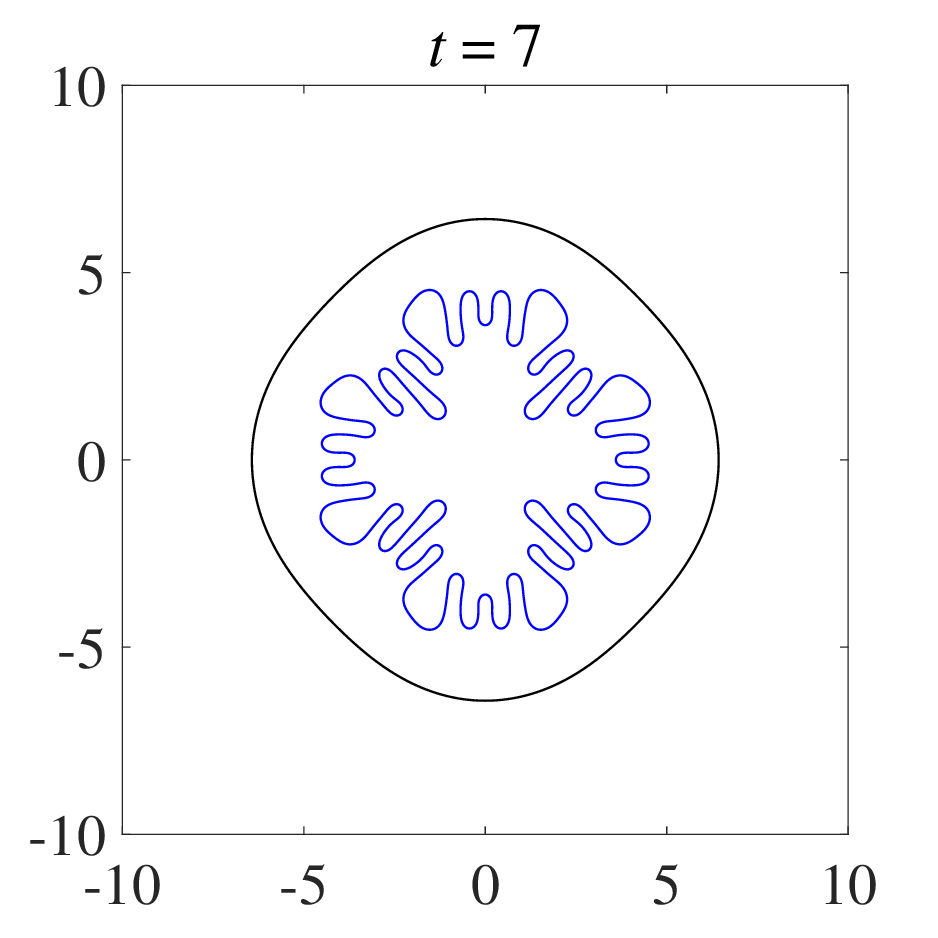}
	\includegraphics[scale=0.27]{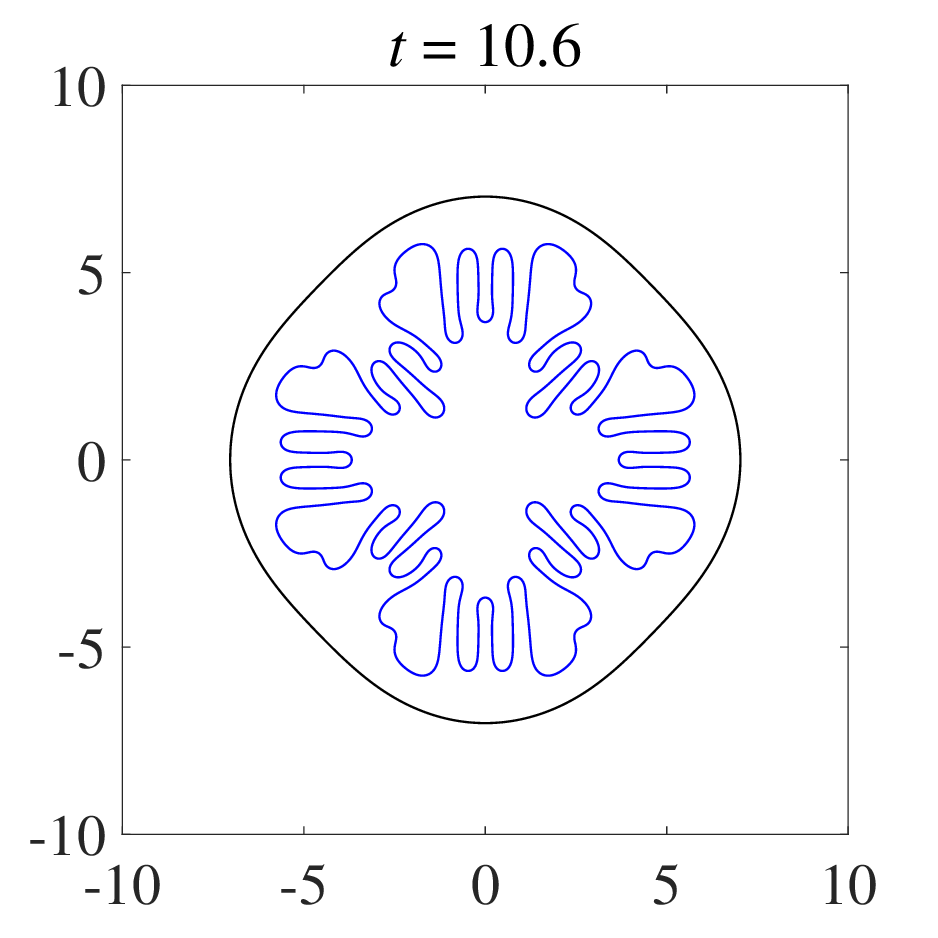}
	\includegraphics[scale=0.27]{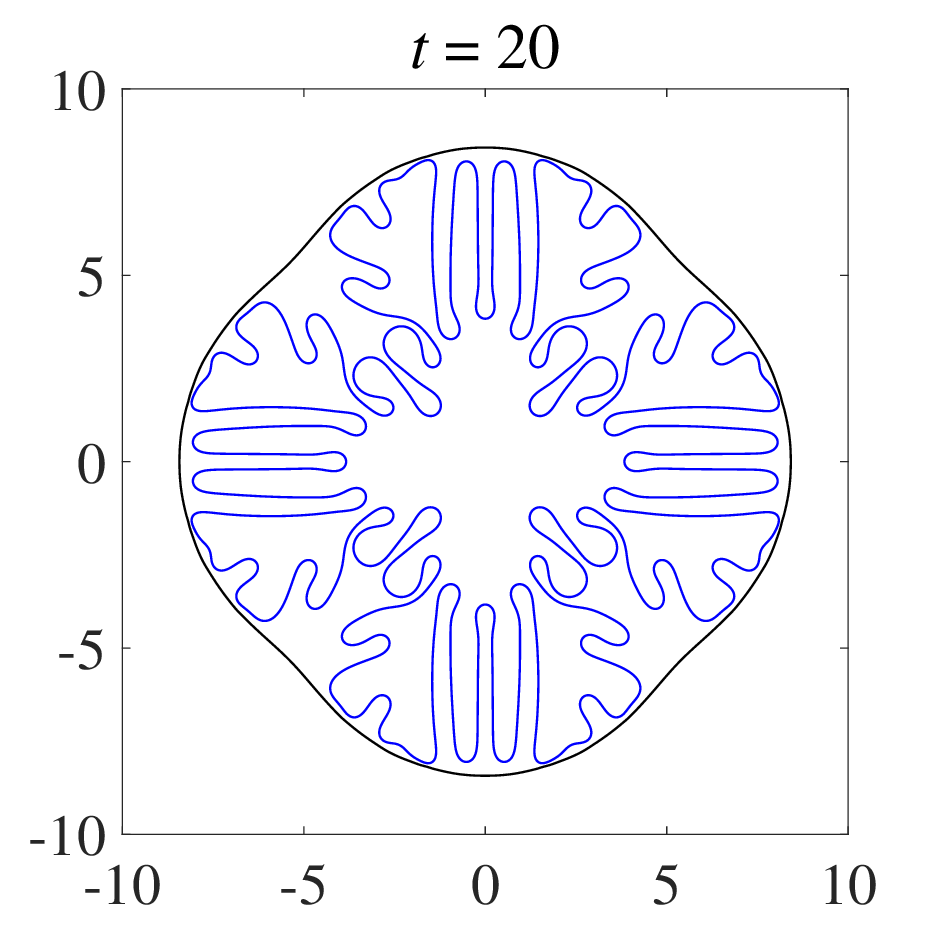}\\
	\includegraphics[scale=0.27]{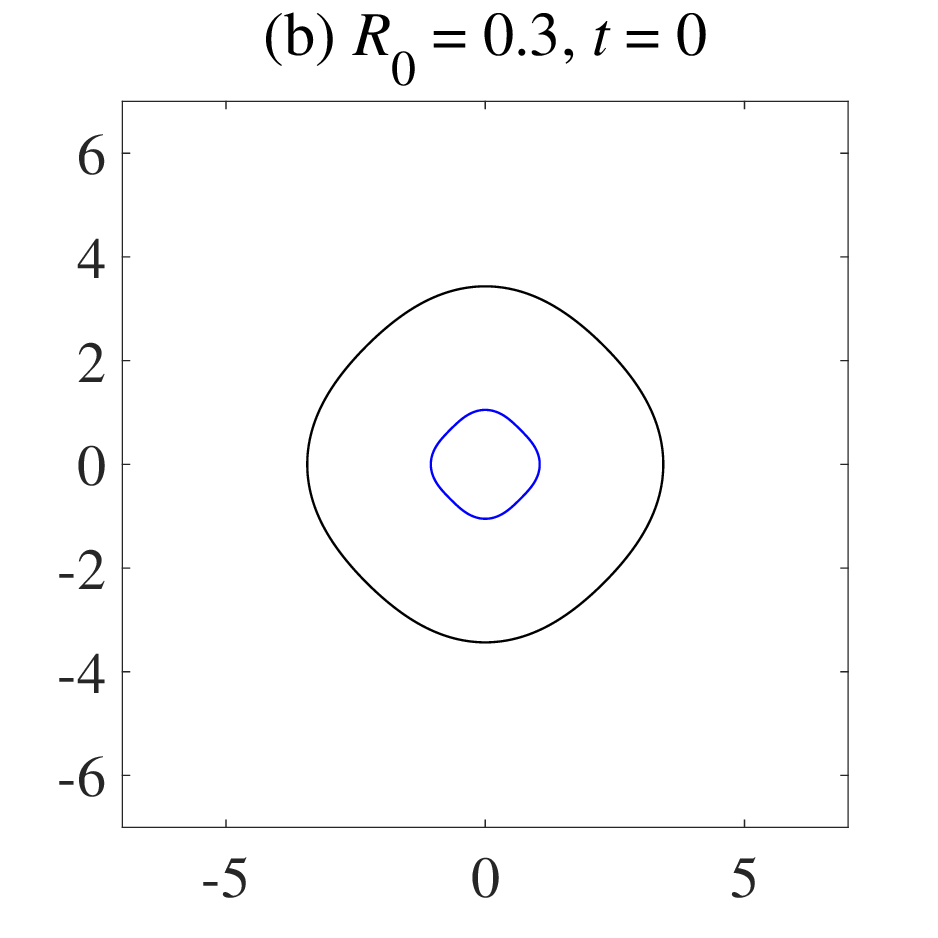}
	\includegraphics[scale=0.27]{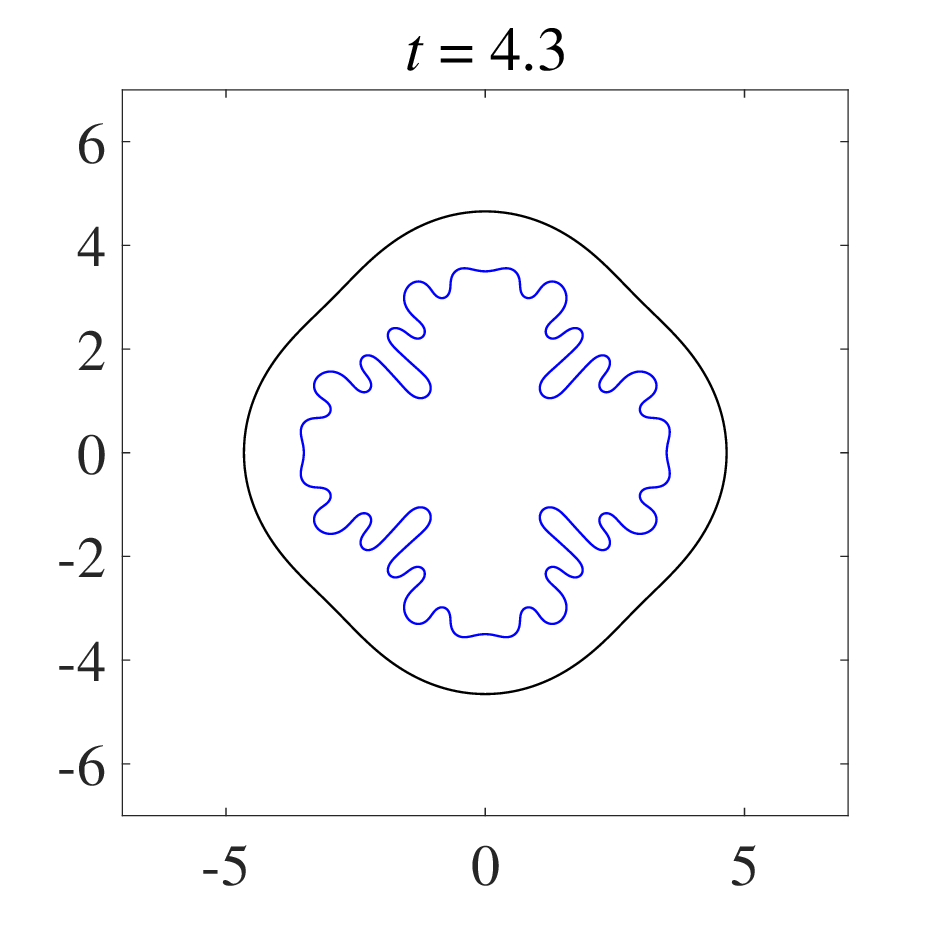}
	\includegraphics[scale=0.27]{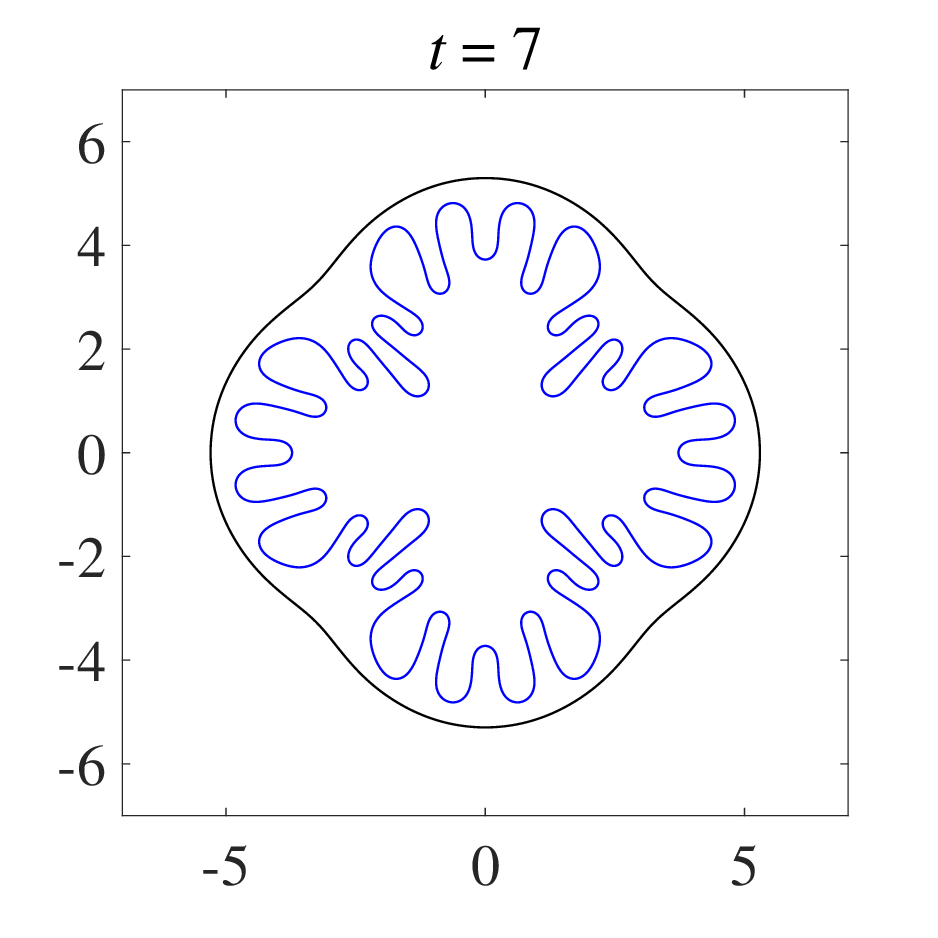}
	\includegraphics[scale=0.27]{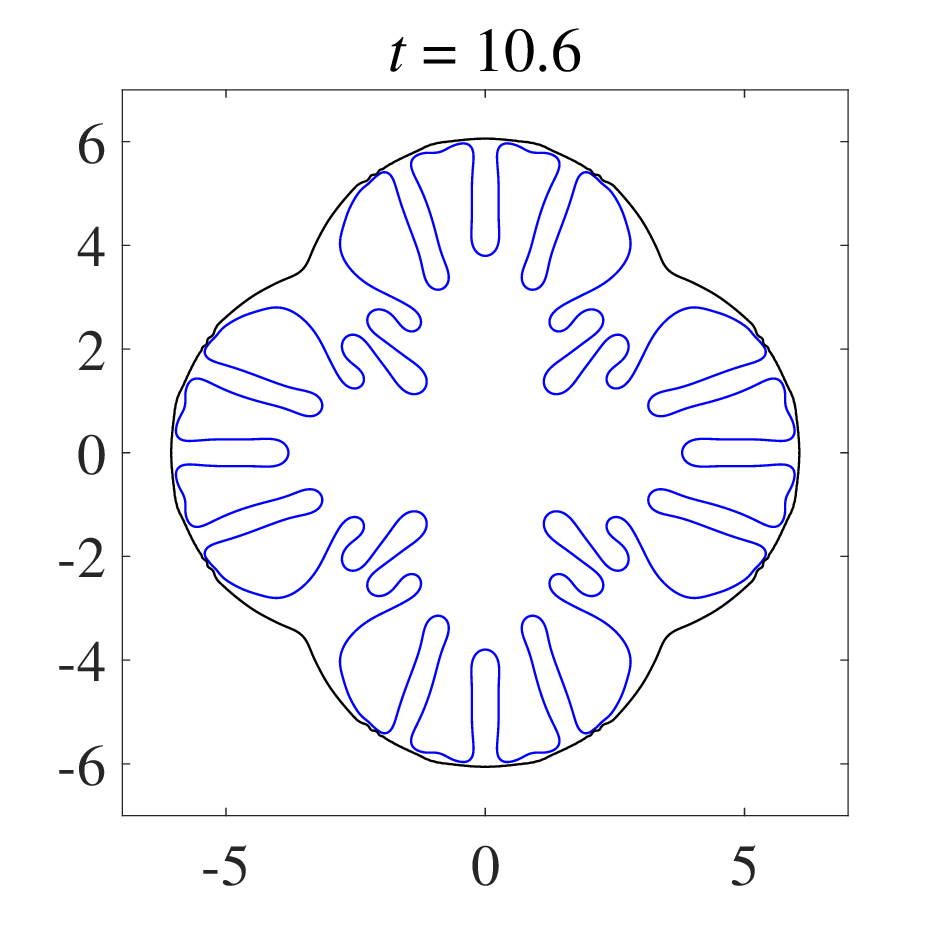}\\
	\includegraphics[scale=0.27]{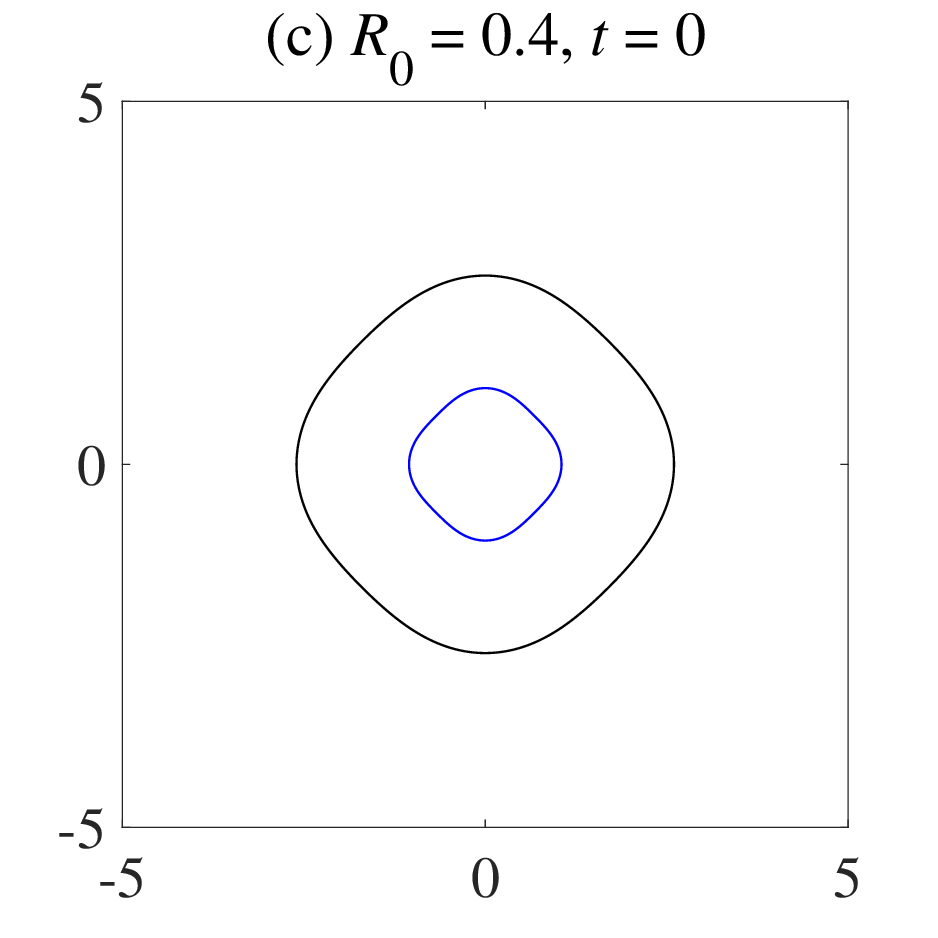}
	\includegraphics[scale=0.27]{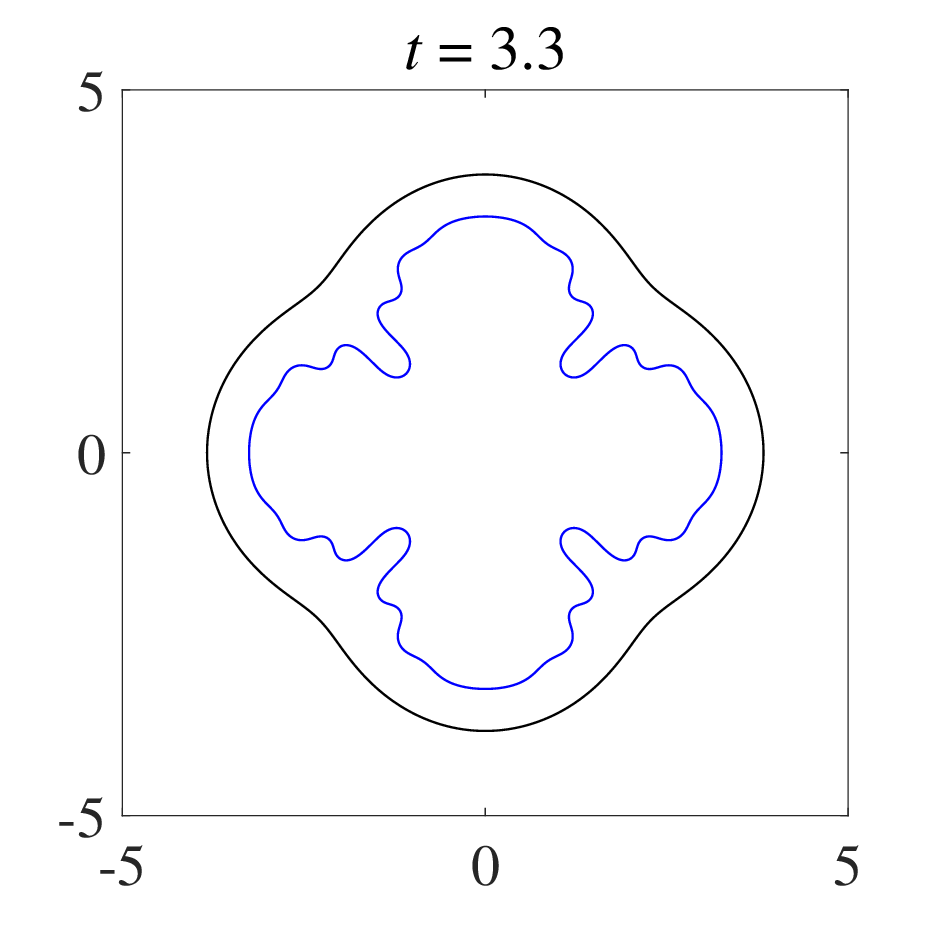}
	\includegraphics[scale=0.27]{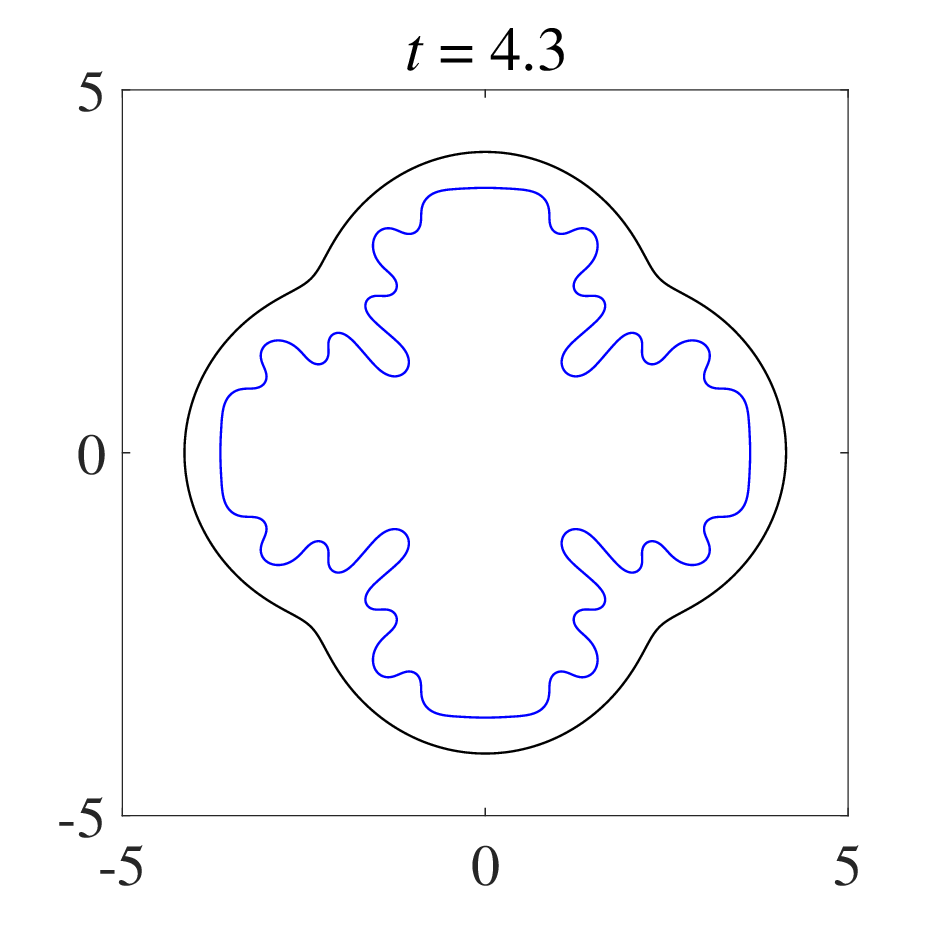}
	\includegraphics[scale=0.27]{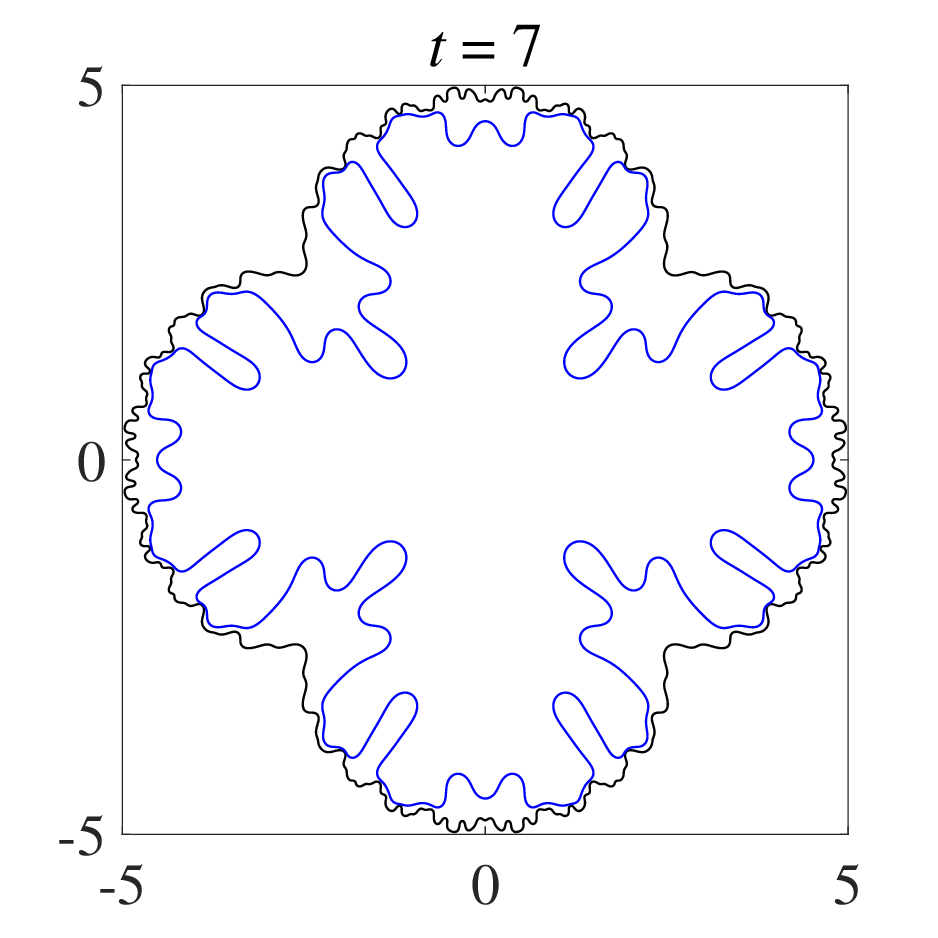}\\	
	\includegraphics[scale=0.27]{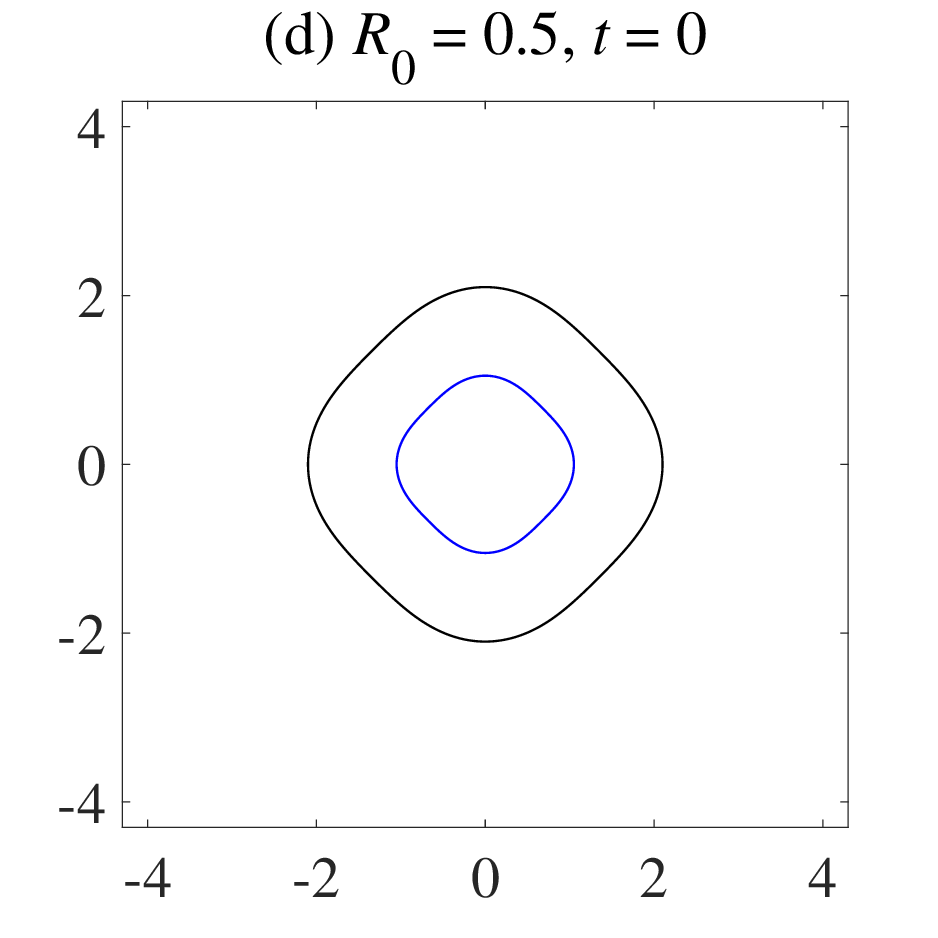}
	\includegraphics[scale=0.27]{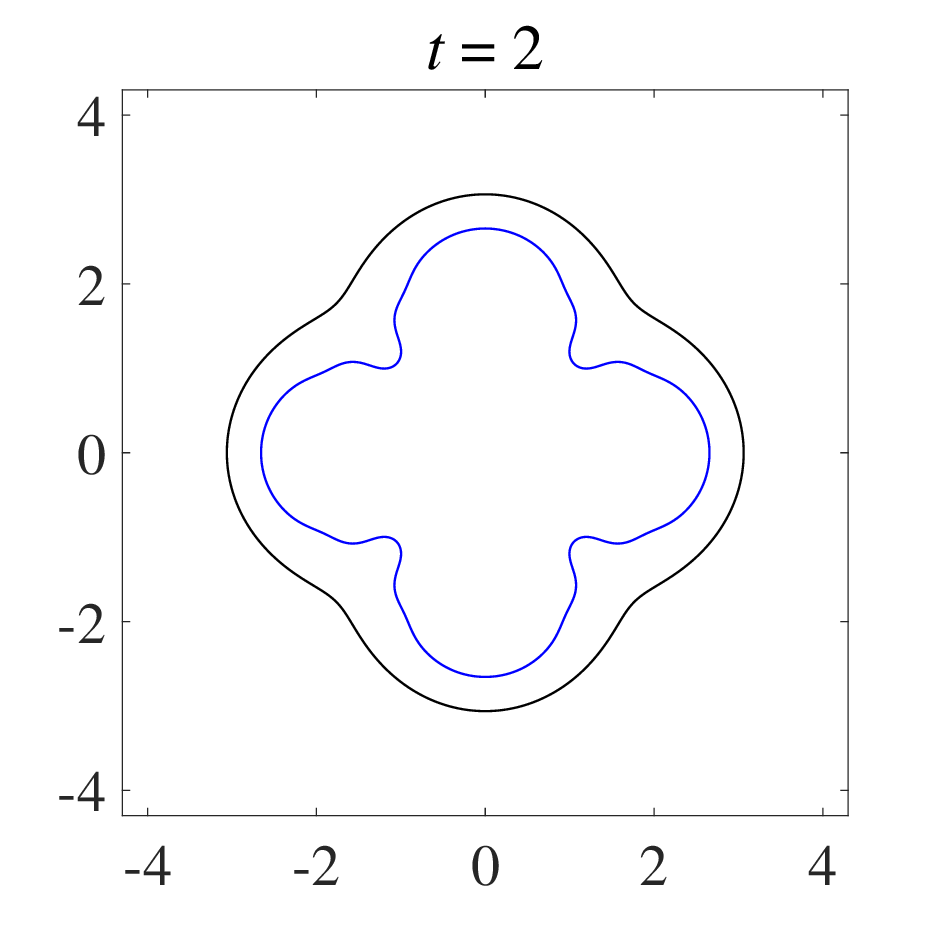}
	\includegraphics[scale=0.27]{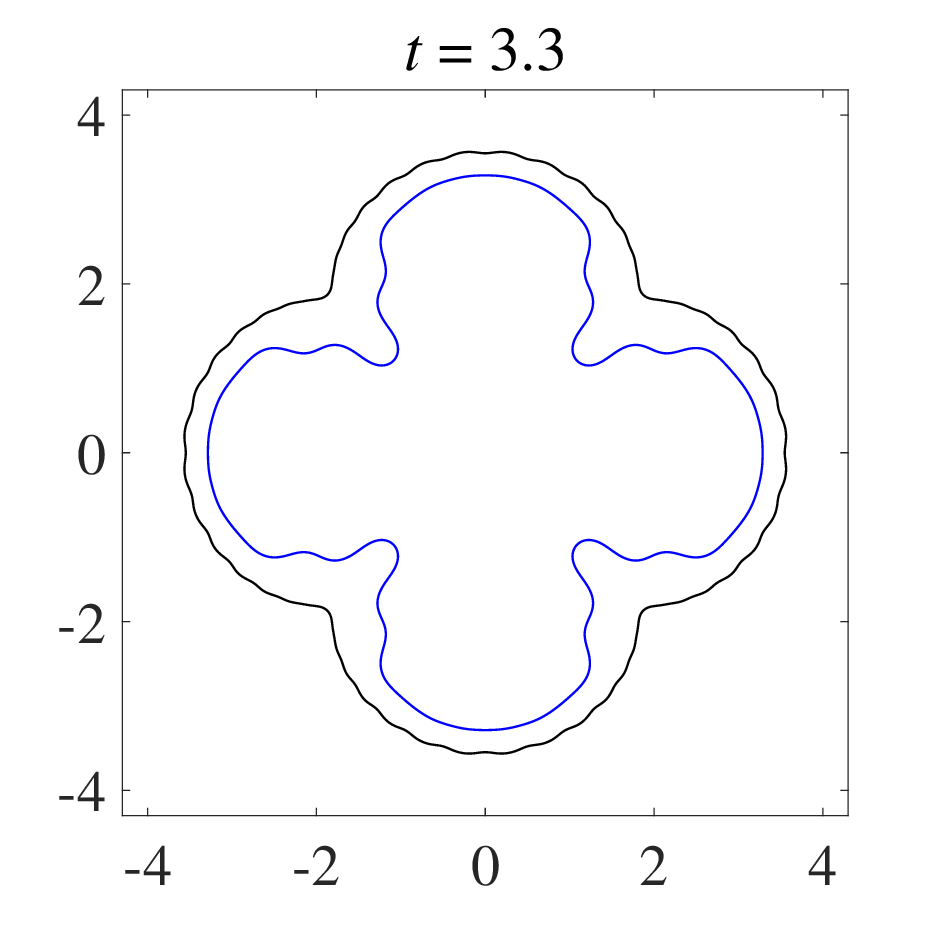}
	\includegraphics[scale=0.27]{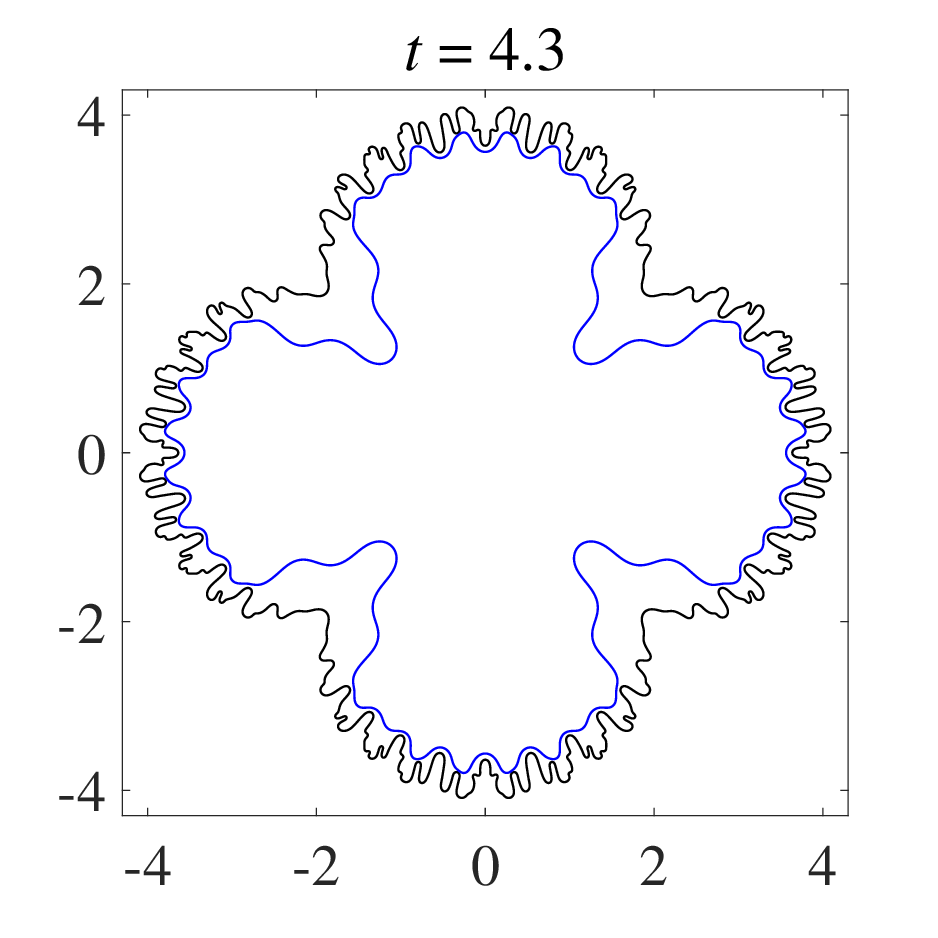}																																
	\caption{Dynamics of the nonlinear interfacial patterns illustrating typical fingering morphologies during three-layer radial Hele-Shaw flows for different initial ratio of the unperturbed radii $R_0$: (a) $R_{0}=0.2$, (b) $R_{0}=0.3$, (c) $R_0=0.4$, and (d) $R_0=0.5$. In a same row, time increases from left to right. In addition, we set ${\rm Ca}=1000$, $\alpha=1$, $\beta_{21}=0.01$, $\beta_{23}=100$, $N=8192$, and $\Delta t=1\times 10^{-4}$.}
	\label{int1}
\end{figure}

We begin our analysis by examining  the impact of the initial distance $d$ between the interfaces on the dynamics of the coupled-interface system. We set the viscosities of the fluids in such a way that both interfaces are unstable and have the same surface tension, namely $\beta_{21}=0.01$, $\beta_{23}=100$, $\alpha=1$, and ${\rm Ca}=1000$. In Fig.~\ref{int1} we illustrate the typical nonlinear morphologies that may emerge during the evolution. Each row of Fig.~\ref{int1} depicts the temporal evolution snapshots of both interfaces for a given value of initial ratio of the unperturbed radii $R_0$: (a) $R_{0}=0.2$, (b) $R_{0}=0.3$, (c) $R_0=0.4$, and (d) $R_0=0.5$. Time increases from left to right and is specified on the top of each frame. The initial shapes for the inner and outer interfaces are given, respectively, by ${\cal R}_{1}(\theta,0)=1+0.05\cos(4\theta)$ and ${\cal R}_{2}(\theta,0)=1/R_0+ 0.1\cos(4\theta)$.  In addition, we set $N=8192$ and $\Delta t=1\times 10^{-4}$. 

Before we advance, we first explain how we determine the final time $t_{f}$. The interfaces shown in Fig.~\ref{int1} for various $R_{0}$ have been obtained after time has evolved in the interval $0 \leqslant t \leqslant t_{f}$. It should be stressed that the values of the final time $t_{f}$ used in Fig.~\ref{int1}, and in other parts of this work are not arbitrary. While plotting the interfaces depicted in this paper, we stop the time evolution of the patterns as soon as the minimum distance between the outer and inner interfaces is only about $6h$, where $h$ is the spatial resolution of the inner interface.

By inspecting Fig.~\ref{int1}(a) with $R_{0}=0.2$, we observe that the initial small fingers grow and spread outward. For the outer interface, we can identify the formation of just a mildly deformed fourfold shape presenting small protuberances of same length. This outer pattern is considerably simpler than the one exhibited by the inner interface, which presents a fully developed, highly branched fingering pattern. In the formation of this convoluted shape, as the fingers advance their tips become wide and flat. At this point, some of these flat-tip fingers will eventually split through finger-tip splitting process. It is precisely the multiple occurrences of this nonlinear pattern-forming mechanism that is responsible for generating such a ramified inner interfacial pattern. Another interesting nonlinear behavior revealed by the inner boundary shape is the larger fingers shielding the growth of the smaller adjacent ones, which is responsible for the intense variability among the lengths of fingers. These findings suggest that our boundary integral method is capable of reproducing the mechanisms of spreading, splitting, and shielding observed in real experiments of the classical two-layer radial Hele-Shaw cell flow~\cite{Lp,mccloud1995experimental}. Therefore, when $R_{0}=0.2$, the coupling between the interfaces is weak and although the inner and outer interfaces are very close together at the final time $t=20$, this is not enough to induce the development of nonlinear ramifications on the outer interface. Moreover, finger competition (related to finger length variability) is intense among the fingers of the inner interface and absent among the outer interface fingers. 

In Fig.~\ref{int1}(b) we consider the value $R_{0}=0.3$. Despite the fact that the outer interface remains almost identical to the previous situation ($R_{0} = 0.2$), perhaps just presenting a slight increase on the amplitude of the fingers, for this higher value of $R_{0}$ one notices an attenuation on the nonlinear ramifications on the inner interface, leading ultimately to the formation of a less branched fingering pattern where the occurrence of finger-tip splitting events is diminished. This attenuation effect provided by the coupling of the interfaces becomes even more evident when a larger value of $R_{0}$ is considered, as in Fig.~\ref{int1}(c) for $R_{0}=0.4$. It is clear that the inner interface is more stable while the outer one now evolves and develops four sizable fingers with small bumps at their tips. These unusual bumps, which are not commonly seen neither in experiments and simulations of classic two-layer radial Hele-Shaw flow, seem to arise as a consequence of the three-layer, double-interface system and are associated to the spontaneous growth of high-frequency modes.

Now, we turn our attention to Fig.~\ref{int1}(d), which considers the value $R_{0}=0.5$. First, both interfaces now evolve into a fourfold-like shape. Unlike the $R_{0}=0.2$ case shown in Fig.~\ref{int1}(a), finger-tip splitting events are absent on both interfaces. However, we note the development of small bumps on the tip of each finger on the outer interface, which triggers the emergence of similar a pattern on the inner interface. Also, there is no clear sign of finger competition for that case. This almost-synchronized evolution implies the strong coupling between the two interfaces, in which the nonlinear pattern-forming mechanisms of finger-tip splitting and finger competition are not detected. It is worthwhile to mention that these numerical results are consistent with the theoretical weakly nonlinear findings of Ref.~\cite{pedronew} regarding the role of the fluid annulus in providing an overall attenuation on the nonlinear pattern-forming mechanism and on regulating the final morphologies of the interfaces. As a last remark about the results depicted in Fig.~\ref{int1}, we stress that the numerical results have been double-checked by multiple refinement studies, where we used a larger number of points $N$ along each interface and reduced time steps $\Delta t$. All these tests resulted in the same interfacial morphologies depicted in Fig.~\ref{int1}, indicating that the bumps are indeed a physical formation.

\begin{figure}[t]
	\includegraphics[scale=0.45]{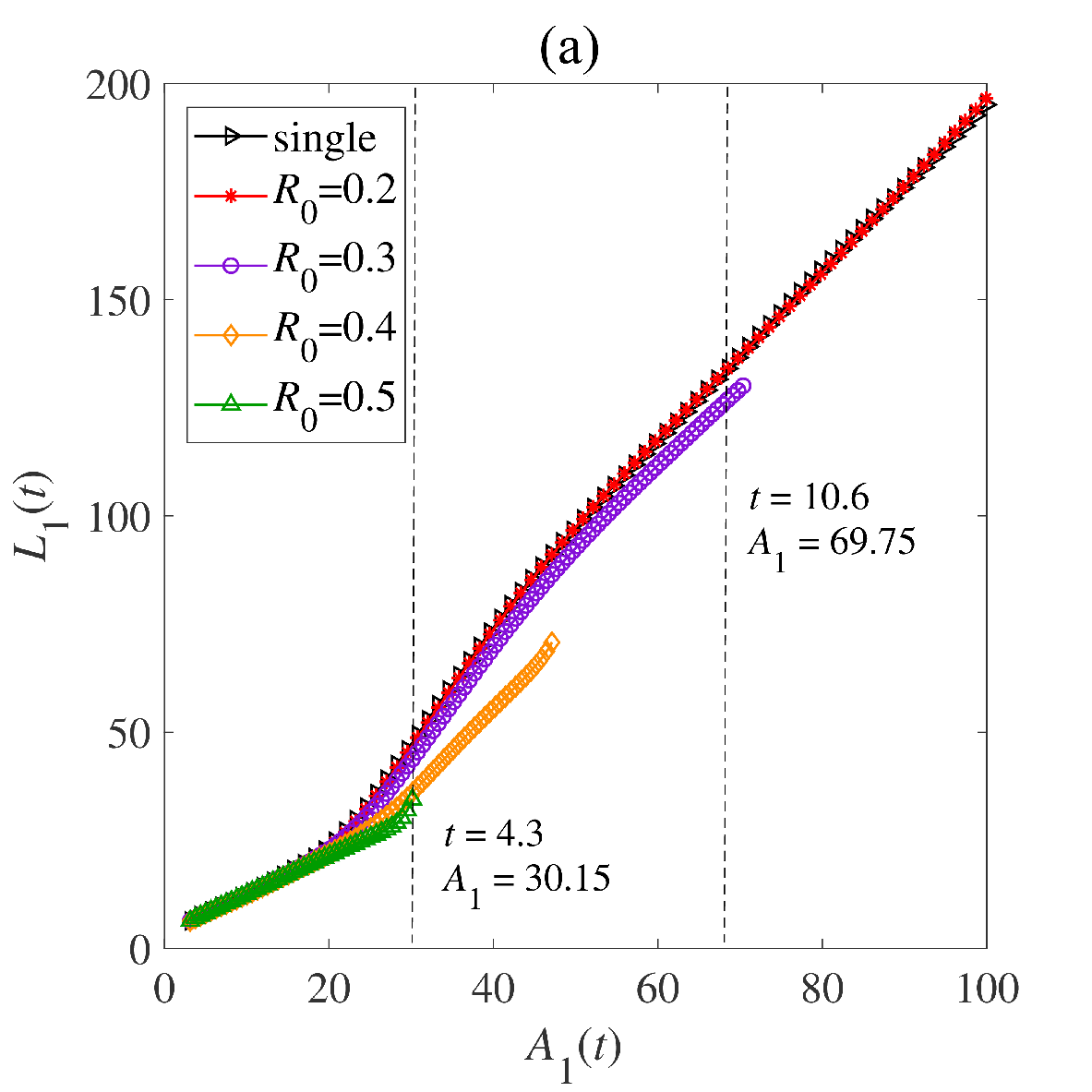}\\
	\includegraphics[scale=0.53]{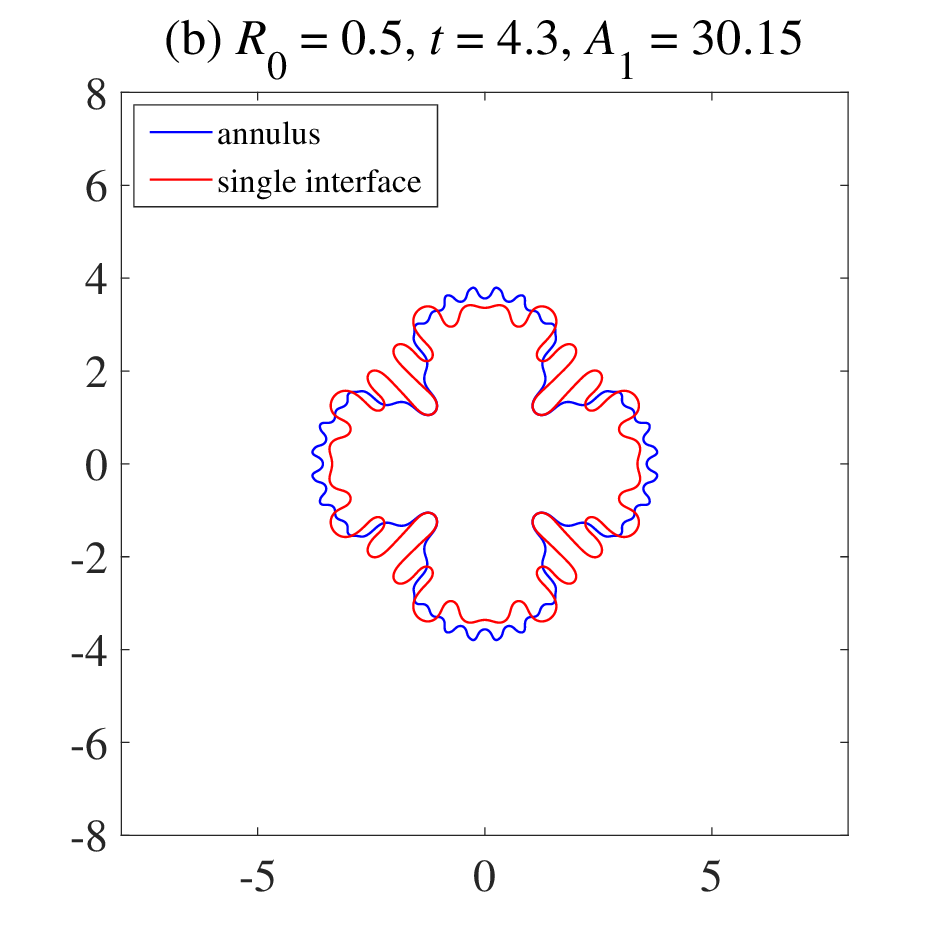}
	\includegraphics[scale=0.53]{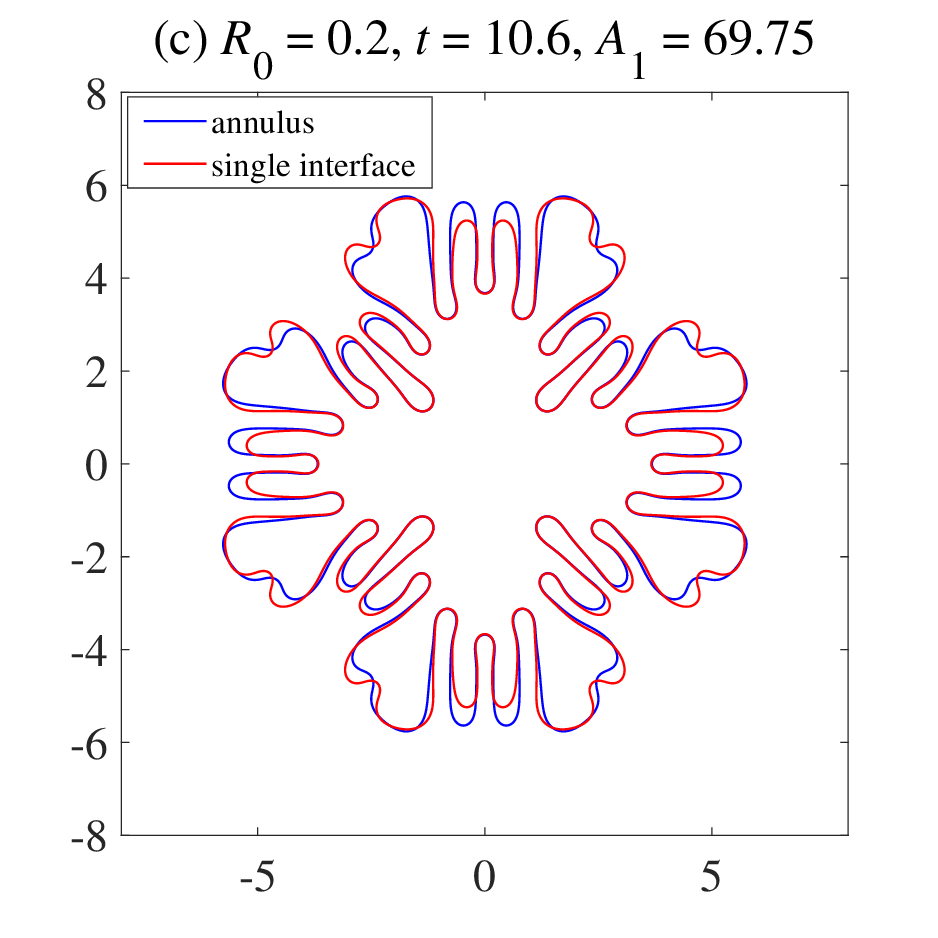}
	\caption{(a) Parametric plot expressing the behavior of the length $L_{1}(t)$ of the inner interface with respect to variations in its area $A_{1}(t)$, for four values of initial ratio of the unperturbed radii $R_{0}$ and for the two-fluid single-interface flow ($R \to 0$). Bottom panels: Comparison between the morphologies of the inner interface generated in the three-layer flow (blue curve) and the corresponding two-layer single-interface flow (red curve). The three-layer morphologies are obtained considering (b) $R_0=0.5$ and $t=4.3$, (c) $R_0=0.2$ and $t=10.6$. Here, the outer interfaces are not displayed. All physical parameters and initial conditions are the same as the ones used in Fig.~\ref{int1}.}
	\label{int1para}
\end{figure}

From the inspection of Fig.~\ref{int1}, it is evident that variations in $R_{0}$ have a great impact on the emerging fingering structures, especially on the final morphology of the inner interface. Although the final shapes presented by the outer interface do not change dramatically at early times, the morphologies acquired by the inner interface are notable different as $R_{0}$ increases, going from a highly unstable branched pattern formed by multiple finger-tip splitting events for lower values of $R_{0}$, to a more stable fourfold structure for larger values of $R_{0}$. To investigate these instability issues and also to provide a quantitative comparison between all the different shapes acquired by the inner interface as $R_{0}$ is varied, in Fig.~\ref{int1para}(a) we present a parametric plot expressing the behavior of the length $L_{1}(t)$ of the inner interface with respect to variations in its area $A_{1}(t)$, for the situations depicted in Fig.~\ref{int1}. The same set of physical parameters and initial conditions for Fig.~\ref{int1} are now used in Fig.~\ref{int1para}. 

The type of graph portrayed in Fig.~\ref{int1para}(a) is convenient to explain the morphologies that arise in Fig.~\ref{int1} for the inner interface for each value of $R_{0}$, since for a given value of $A_{1}(t)$ the most unstable situation is the one related to the largest value of $L_{1}(t)$. 
Besides the four values of $R_{0}$ used in Fig.~\ref{int1}, in Fig.~\ref{int1para}(a) we have also included an extra situation related to the usual two-fluid single-interface version of the flow, i.e., the radial displacement of fluid 2 by fluid 1 without the presence of the third layer of fluid 3. This is done to completely remove the effects related to the coupling between the interfaces and it is equivalent to take the limit $R \to 0$ in our equations.

\begin{figure}[t]
	\includegraphics[scale=0.45]{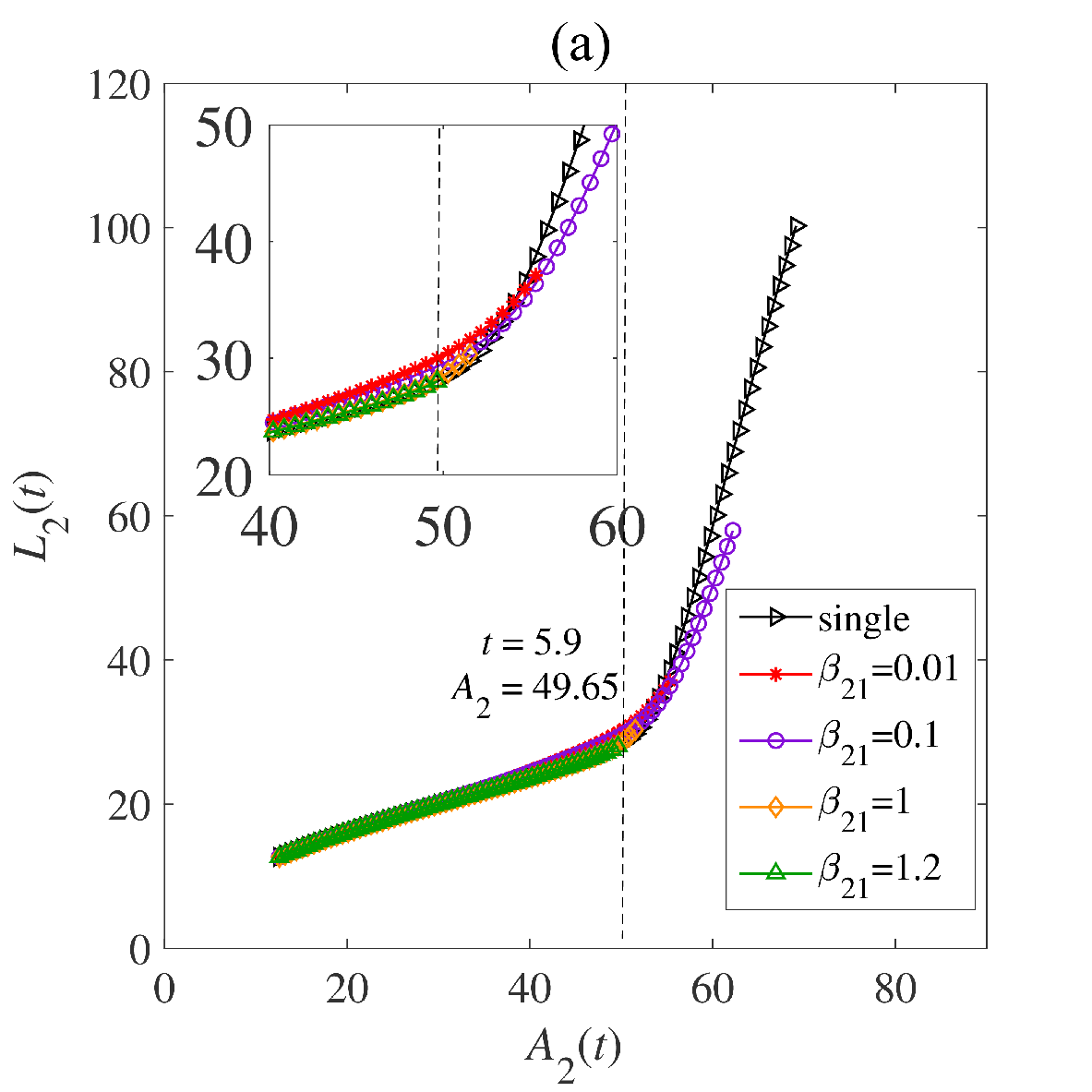}\\
	\includegraphics[scale=0.53]{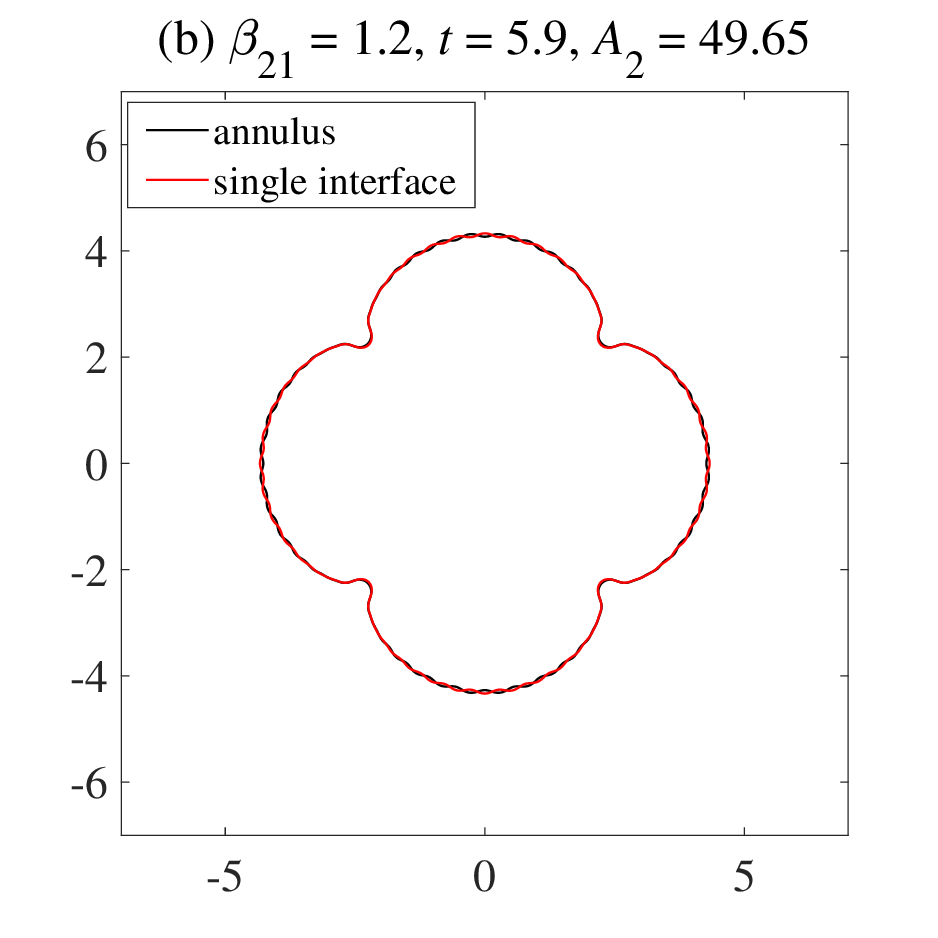}
	\includegraphics[scale=0.53]{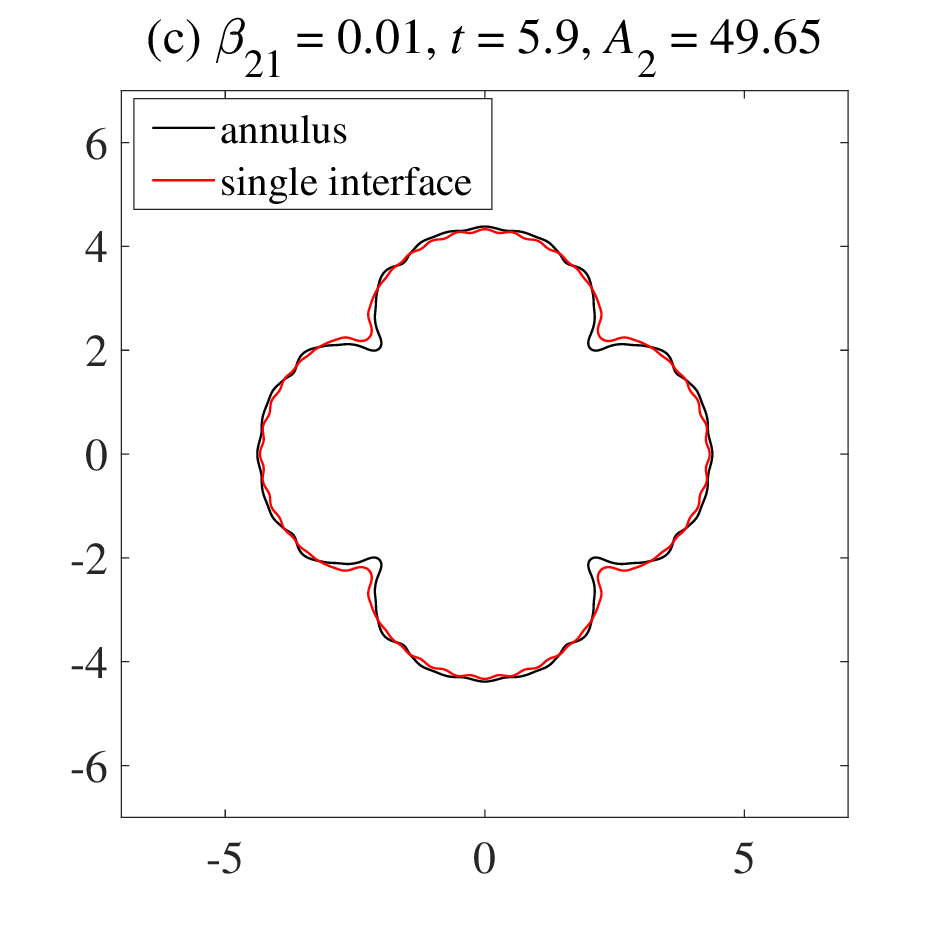}
	\caption{(a) Parametric plot expressing the behavior of the length $L_{2}(t)$ of the outer interface with respect to variations in its area $A_{2}(t)$. We set $\beta_{23}=10$ and use four different values for $\beta_{21}$: $0.01$, $0.1$, $1$, and $1.2$. The two-fluid single outer interface case is included as well. In addition, we consider $R_{0}=0.5$, ${\rm Ca}=1000$, and $\alpha=1$. Bottom panels: Comparison between the morphologies of the outer interface generated in the three-layer flow (black curve) and the corresponding two-layer single-interface flow (red curve). The three-layer morphologies are obtained considering $t=5.9$ and (b) $\beta_{21}=1.2$, and (c) $\beta_{21}=0.01$. Here, the inner interfaces are not displayed.}
	\label{para12}
\end{figure}

By examining Fig.~\ref{int1para}(a), it is apparent that, at early times of the dynamics, all the curves are superposed regardless of the value of $R_{0}$. This observation suggests that initially, the impact of the fluid annulus on the inner interface is very small. However, as time increases, the curves separate from each other and the effects of the fluid annulus become sizable. Moreover, it is clear that all the three curves related to the three-layer flow with $R_{0} > 0.2$ are below the two-layer single-interface curve, indicating an evident attenuation of the inner interface instabilities for larger values of $R_{0}$. A comparison between the inner interface formed in the three-layer system for $R_{0}=0.5$ and the corresponding single-interface case is provided in Fig.~\ref{int1para}(b), where the final time is $t=4.3$. As predicted by Fig.~\ref{int1para}(a), the two-layer single-interface seems indeed more unstable than the three-layer inner interface by exhibiting developed long fingers and finger-tip splitting events. On the other hand, the curve for $R_{0}=0.2$ is almost identical to the single-interface curve for the values of time considered here, and one can conclude that for the situation of a thick annulus the evolution of the inner interface in the three-layer system is very similar to the usual two-layer flow~\cite{bataille1968stabilite,Wilson,Lp,Rauseo,May,chen1987radial,chen1989growth,HMVY,OH,PM,mathiesen2006universality,ShuwangPRL,Zhao17}. This fact can be confirmed in Fig.~\ref{int1para}(c), where we compare the inner interface morphology generated in the three-layer system for $R_{0}=0.2$ and the corresponding single-interface case. For the final time $t=10.6$ they are very similar to each other and it is difficult to identify which one is more unstable. The observations extracted
from Fig.~\ref{int1para} are consistent with the equivalent interfacial pattern behaviors depicted in Fig.~\ref{int1} and quantitatively support our numerical simulations. 

\subsubsection{The roles of $\beta_{21}$, $\beta_{23}$, and $\alpha$}
\label{beta}

It is a well-known fact that the viscous fingering structures formed at the advance-time regime of the radial Hele-Shaw cell dynamics are strongly influenced by the viscosity ratio $\beta$ between the fluids involved. It has been shown that a large variability of interfacial morphologies is found where distinctively different patterns are created as the viscosity ratio is varied. As reported in Refs.~\cite{bischofberger2015island,perugini2005viscous} and consistently with previous experimental results~\cite{Lp,Rauseo,May,chen1987radial,chen1989growth,HMVY,OH}, the traditional large-viscosity-ratio patterns present long fingers growing from a small circular region in which the outer fluid is completely displaced. Nevertheless, as the viscosity ratio is decreased, considerably dissimilar fingering shapes arise where the mentioned inner circular region increases dramatically while the lengths of the growing fingers tend to decrease. Therefore, in the framework of the usual two-layer radial Hele-Shaw cell flow, larger viscosity ratios are associated with more unstable complex structures, while lower values of viscosity ratios generate less unstable patterns presenting small fingers. However, these behaviors cannot be taken for granted in the three-layer, double-interface system. Changes in any fluids' viscosities can affect both interfaces in a nontrivial way. For instance, varying the viscosity $\mu_{1}$ of fluid 1 while keeping all the other viscosities unchanged will directly impact the development of the inner interface. But since the inner and outer interfaces are coupled, this change in $\mu_{1}$ can also indirectly influence the growth of instabilities on the outer interface.

\begin{figure}[t]
	\includegraphics[scale=0.45]{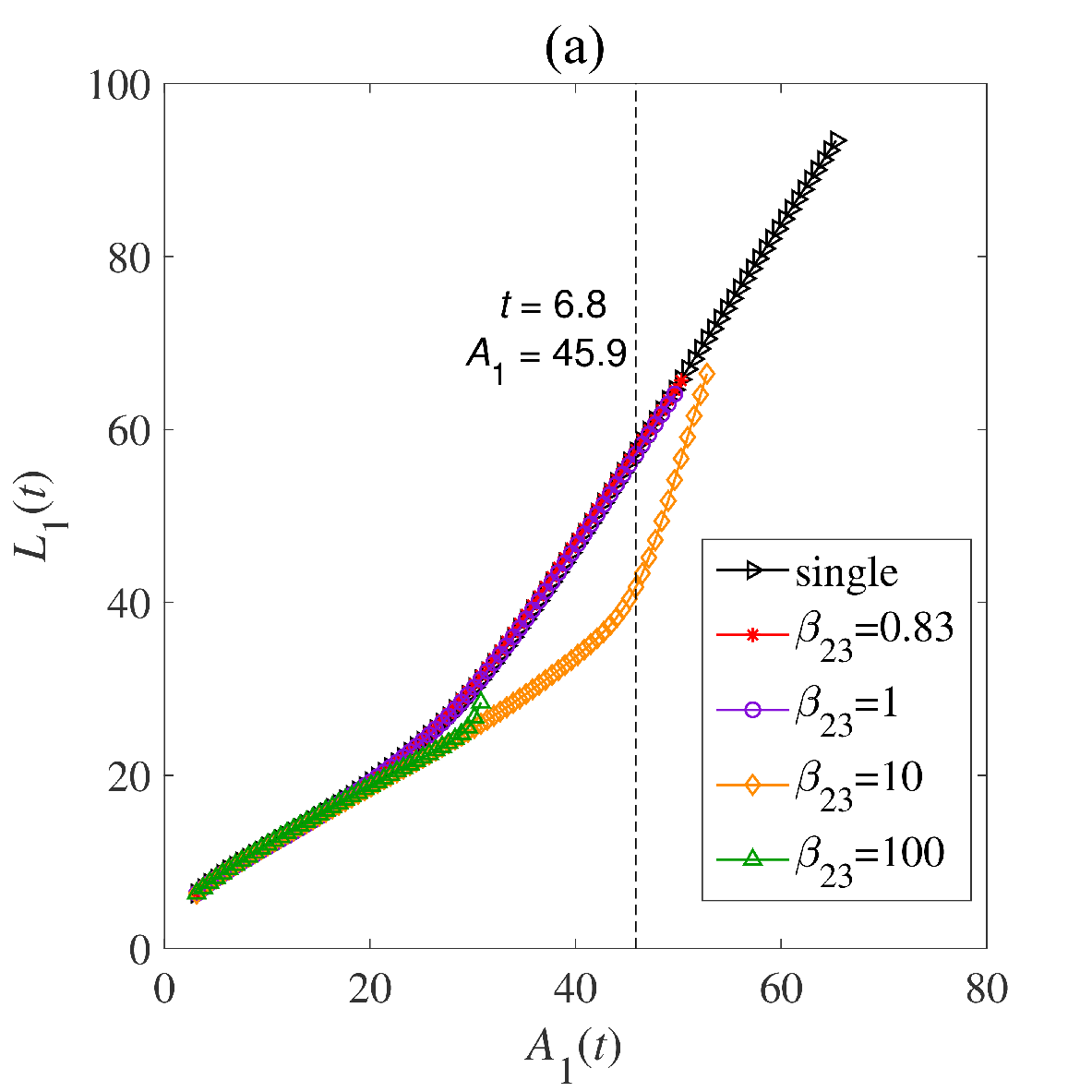}\\
	\includegraphics[scale=0.53]{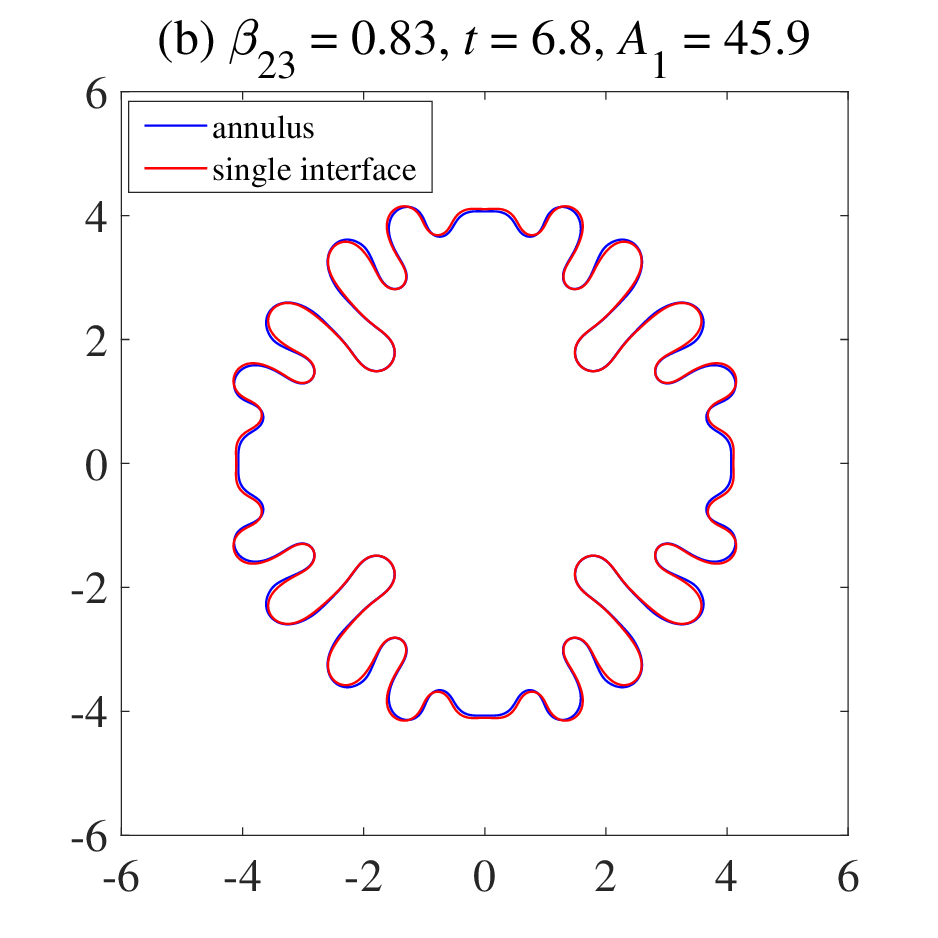}
	\includegraphics[scale=0.53]{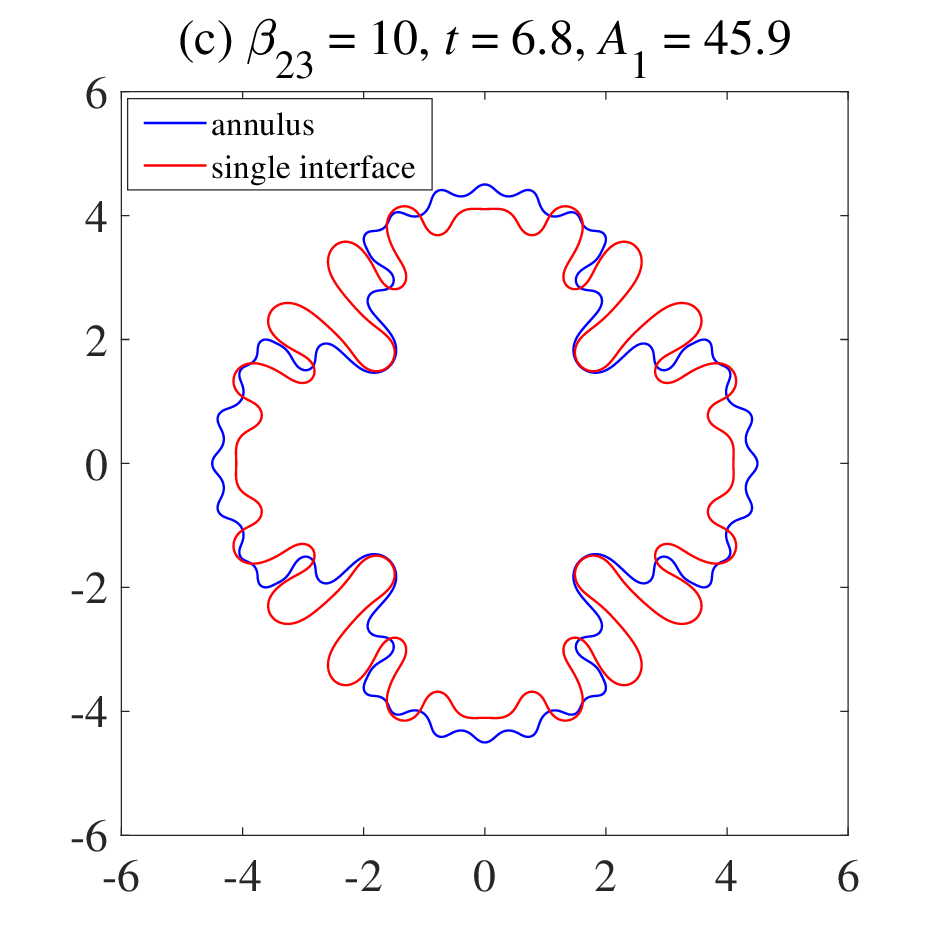}
	\caption{(a) Parametric plot expressing the behavior of the length $L_{1}(t)$ of the inner  interface with respect to variations in its area $A_{1}(t)$. We set $\beta_{21}=0.1$ and use four different values for $\beta_{23}$: $0.83$, $1$, $10$, and $100$. The two-fluid single inner interface case is included as well. In addition, we consider $R_{0}=0.5$, ${\rm Ca}=1000$, and $\alpha=1$. Bottom panels: Comparison between the morphologies of the inner interface generated in the three-layer flow (blue curve) and the corresponding two-layer single-interface flow (red curve). The three-layer morphologies are obtained considering $t=6.8$ and (b) $\beta_{23}=0.83$, and (c) $\beta_{23}=10$. Here, the outer interfaces are not displayed.}
	\label{para122}
\end{figure}

To probe the impact of the fluids' viscosities on the coupled interfaces, we plot Figs.~\ref{para12} and~\ref{para122}, in which the initial shapes for the inner and outer interfaces are given, respectively, by ${\cal R}_{1}(\theta,0)=1+0.05\cos(4\theta)$ and ${\cal R}_{2}(\theta,0)=1/R_0+ 0.1\cos(4\theta)$,  where $R_{0}=0.5$. The physical parameters considered are $\alpha=1$ and ${\rm Ca}=1000$. Fig.~\ref{para12}(a) represents the situation in which the viscosity $\mu_{1}$ of fluid 1 is varied while $\mu_{2}$ and $\mu_{3}$ remain constant. In our problem, this is equivalent to vary the viscosity ratio $\beta_{21}$ keeping $\beta_{23}$ unchanged. We investigate the indirect impact of fluid 1 on the outer interface by showing a parametric plot expressing the behavior of the length $L_{2}(t)$ of the outer interface with respect to variations in its area $A_{2}(t)$ as time advances. In a very similar way, in Fig.~\ref{para122}(a) we present a parametric plot expressing the behavior of the length $L_{1}(t)$ of the inner interface with respect to variations in its area $A_{1}(t)$, for the situation in which the viscosity $\mu_{3}$ of fluid 3 is varied while $\mu_{1}$ and $\mu_{2}$ remain constant. This can be accomplished by varying the viscosity ratio $\beta_{23}$ keeping $\beta_{21}$ unchanged.

In Fig.~\ref{para12}(a) we set $\beta_{23}=10$ and use four different values for $\beta_{21}$: $0.01$, $0.1$, $1$, and $1.2$. Note that the cases for $\beta_{21}\geq1$ are related to an inner interface originally stable since $\mu_{1}\geq\mu_{2}$, while the situations with $\beta_{21}<1$  correspond to an inner interface originally unstable since $\mu_{1}<\mu_{2}$. By inspecting Fig.~\ref{para12}(a), we verify that the two curves with $\beta_{21}\geq1$ are superposed with the curve corresponding to the single outer interface flow (i.e., radial displacement of fluid 3 by fluid 2 without the presence of fluid 1), regardless of the value of time considered. Therefore, when one considers an originally stable inner interface the impact on the dynamics of the outer interface is very small, even though the interfaces are strongly coupled. This fact can be confirmed in Fig.~\ref{para12}(b) by contrasting the outer interface formed in the three-layer system for $\beta_{21}=1.2$ and the corresponding single-interface case, both at time $t=5.9$. In this scenario, the inner stable interface evolves as a nearly circular shape while the outer unstable interface develops viscous fingering instabilities. As time progresses, the base of fingers of the outer leading interface merges with the circular inner interface. This is actually in accordance with the results previously obtained in Ref.~\cite{cardoso1995formation}. It has been shown experimentally by Cardoso and Woods that, for subsequent times, the intermediate fluid annulus break up into drops (see their Fig. 19) and eventually separate due to the action of the surface tension, which tends to smooth point edges at the rear at the moment of rupture. This formation of drops has also been observed theoretically by Anjos and Li in Ref.~\cite{pedronew}.

Nonetheless, a notably distinct behavior is found when we turn our attention to the two curves with $\beta_{21}<1$ and the single outer interface curve. At early times, the curves for $\beta_{21}=0.01$ and $0.1$ are above the single outer interface curve. This result suggests that when one considers an inner interface originally unstable, there is an enhancement on the instabilities developed by the outer interface due to the coupling provided by the annulus. However, as time increases the position of the curves changes, and $\beta_{21}<1$ curves get below the single outer interface curve. Therefore, we observe a stabilization effect that results from the thinning of the intermediate fluid annulus as the interfaces approach one another. This leads to an attenuation of the instabilities developed by the outer interface and makes it less unstable when compared to the single outer interface flow, where the mentioned effect is absent. In Fig.~\ref{para12}(c) we provide a comparison between the three-layer outer interface for $\beta_{21}=0.01$ and the corresponding single-interface case, both at time $t=5.9$.

Following a similar approach utilized in Fig.~\ref{para12}(a), we now study the responses of the inner interface to changes in the fluids' viscosities. In Fig.~\ref{para122}(a) we set $\beta_{21}=0.1$ and use four different values for $\beta_{23}$: $0.83$, $1$, $10$, and $100$. The cases for $\beta_{23}\leq1$ ($\beta_{23}>1$) are related to an outer interface originally stable (unstable) since $\mu_{2}\geq\mu_{3}$ ($\mu_{2}<\mu_{3}$). Regardless of the value of time, we observe only one type of behavior: the curves with $\beta_{23}\leq1$ are superposed to the single inner interface curve while curves with $\beta_{23}>1$ tend to stay below the single inner interface case. This demonstrates that when the outer interface is originally stable the fluid annulus does not affect the dynamics of the inner interface significantly, as one can see by comparing the interfaces in Fig.~\ref{para122}(b) for $\beta_{23}=0.83$ and $t=6.8$. However, an originally unstable outer interface impacts the development of the inner interface by turning it less unstable, as indicated by the attenuation of long fingers in Fig.~\ref{para122}(c) for $\beta_{23}=10$ and $t=6.8$.

We have also studied the impact of the ratio of surface tensions $\alpha$ on the development of instabilities on both interfaces during the three-layer radial Hele-Shaw cell flow. Following the same type of analysis conducted in Figs.~\ref{int1para},~\ref{para12}, and~\ref{para122}, we summarize our findings as follows: larger values of $\alpha$ tend to stabilize the development of instabilities on both inner and outer interfaces. In addition, this effect is significantly more evident at later times of the dynamics.

\subsubsection{Evolution of the interfaces for arbitrary initial conditions}
\label{initial}

\begin{figure}[t]
	\includegraphics[scale=0.27]{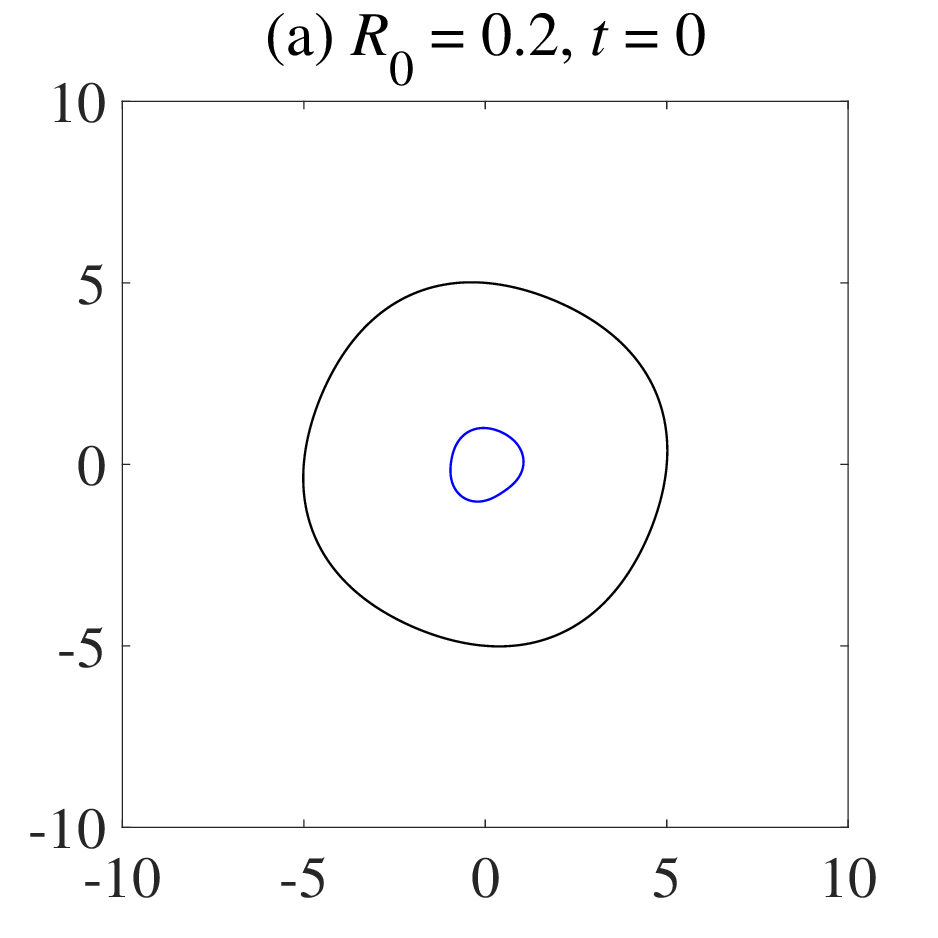}
	\includegraphics[scale=0.27]{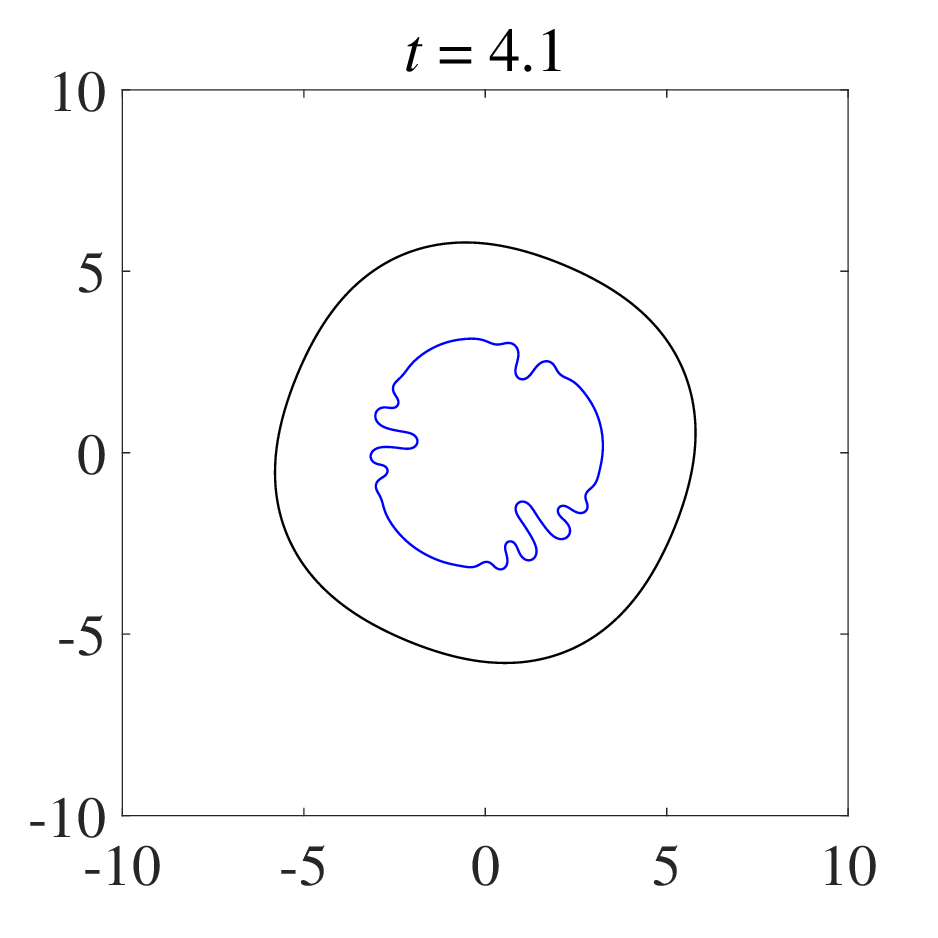}
	\includegraphics[scale=0.27]{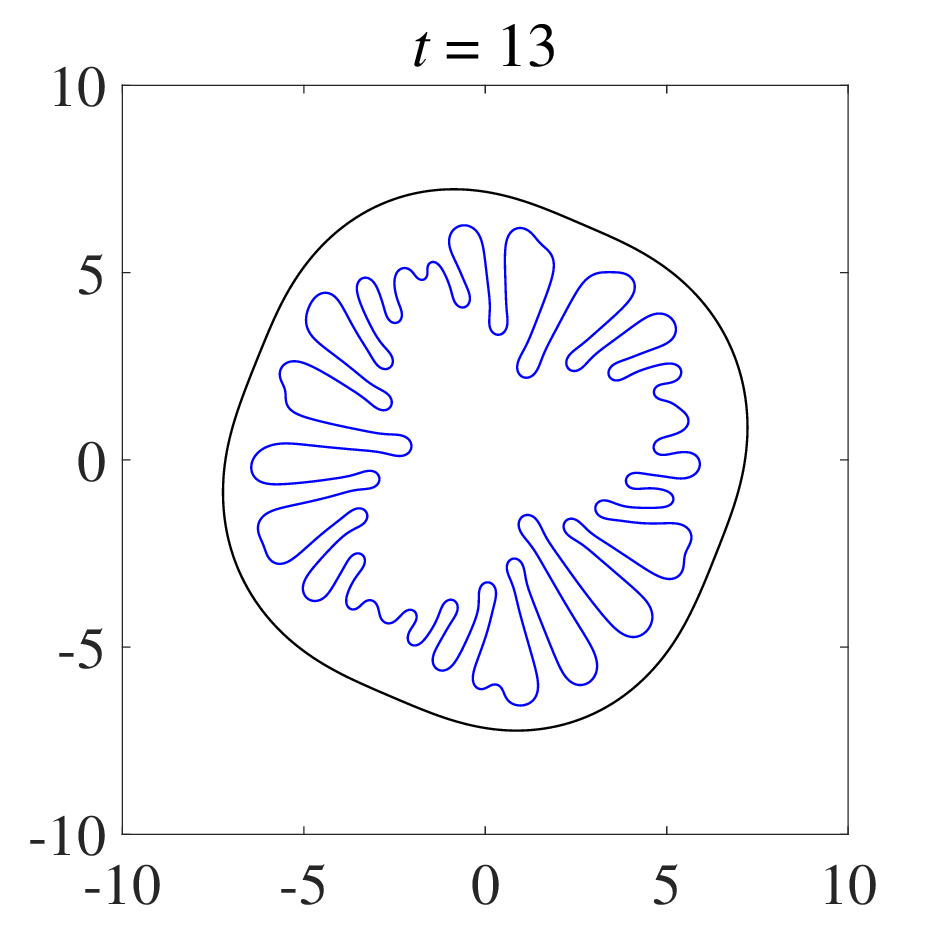}
	\includegraphics[scale=0.27]{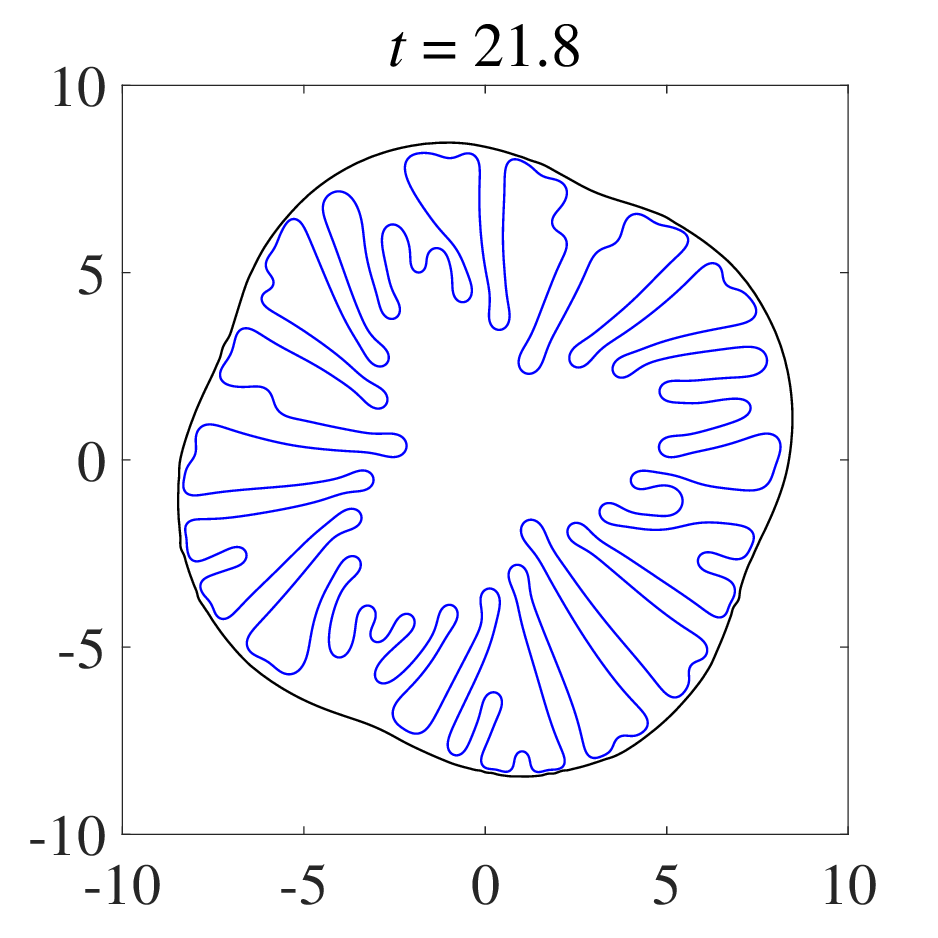}\\
	\includegraphics[scale=0.27]{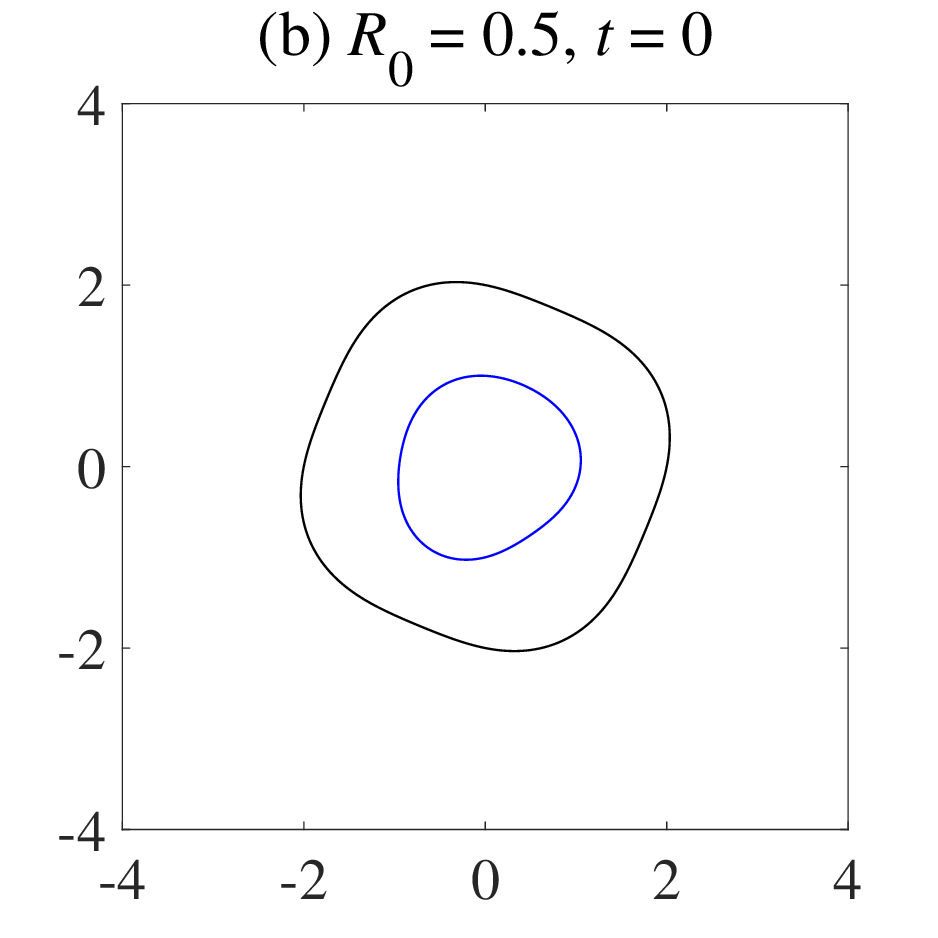}
	\includegraphics[scale=0.27]{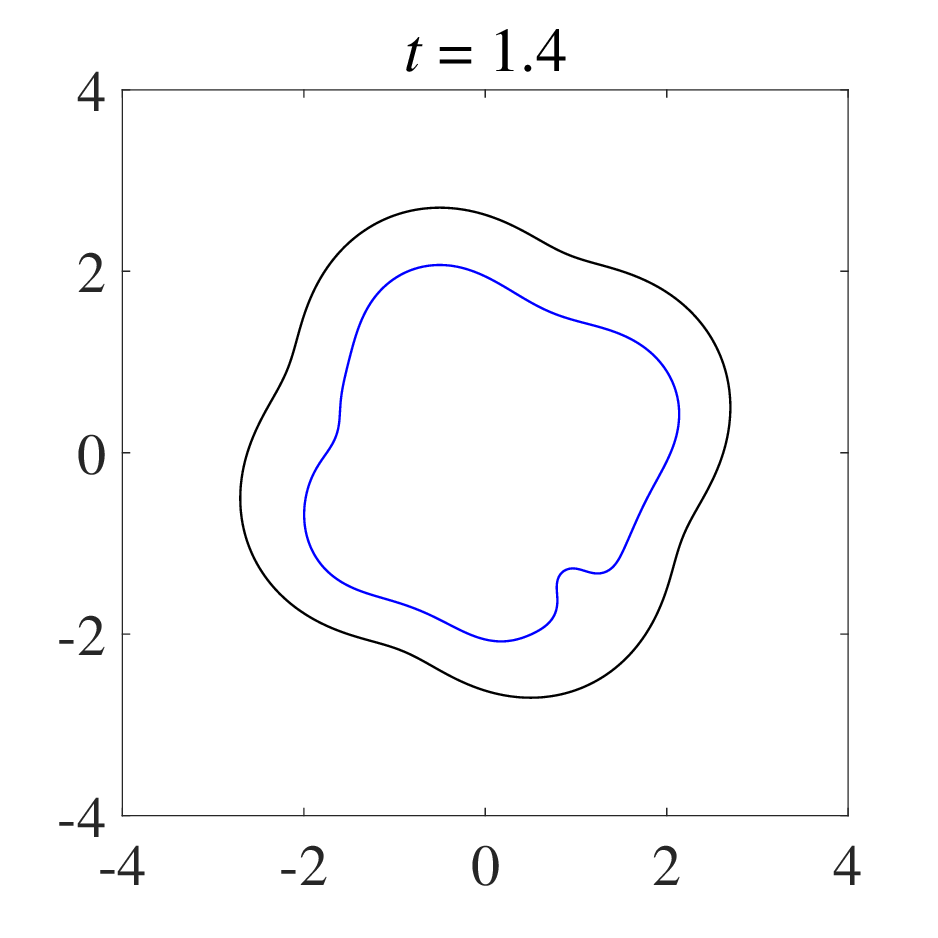}
	\includegraphics[scale=0.27]{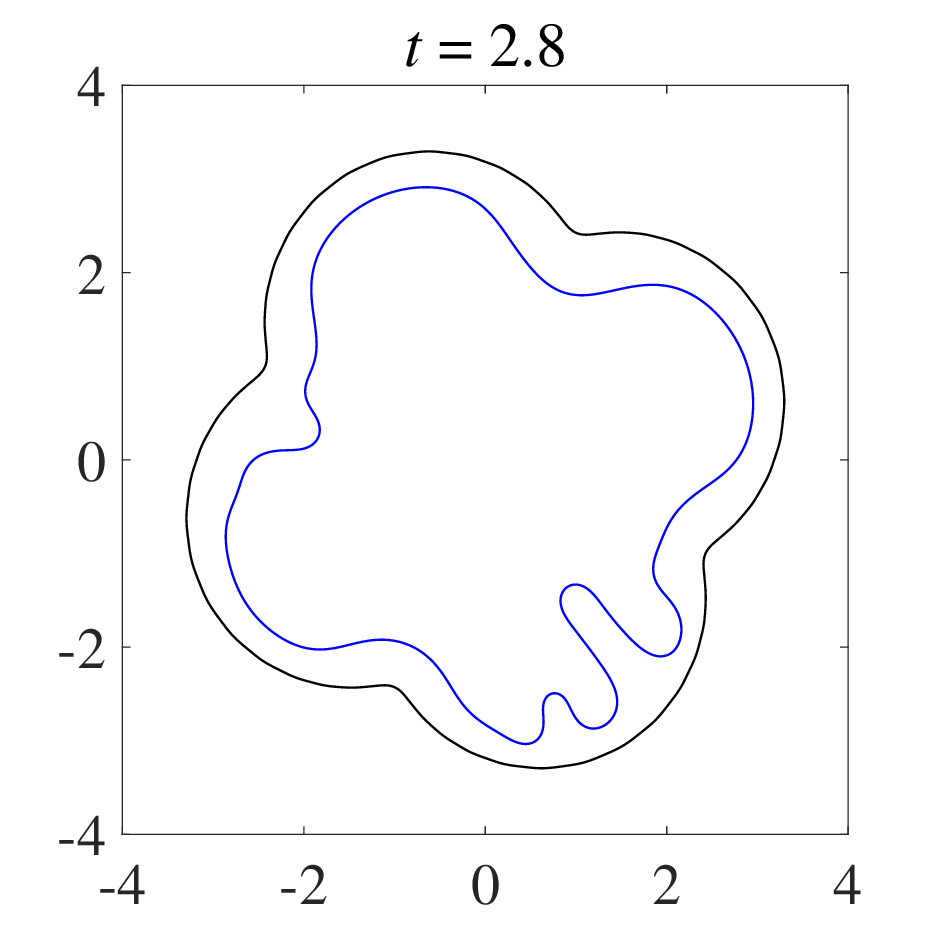}
	\includegraphics[scale=0.27]{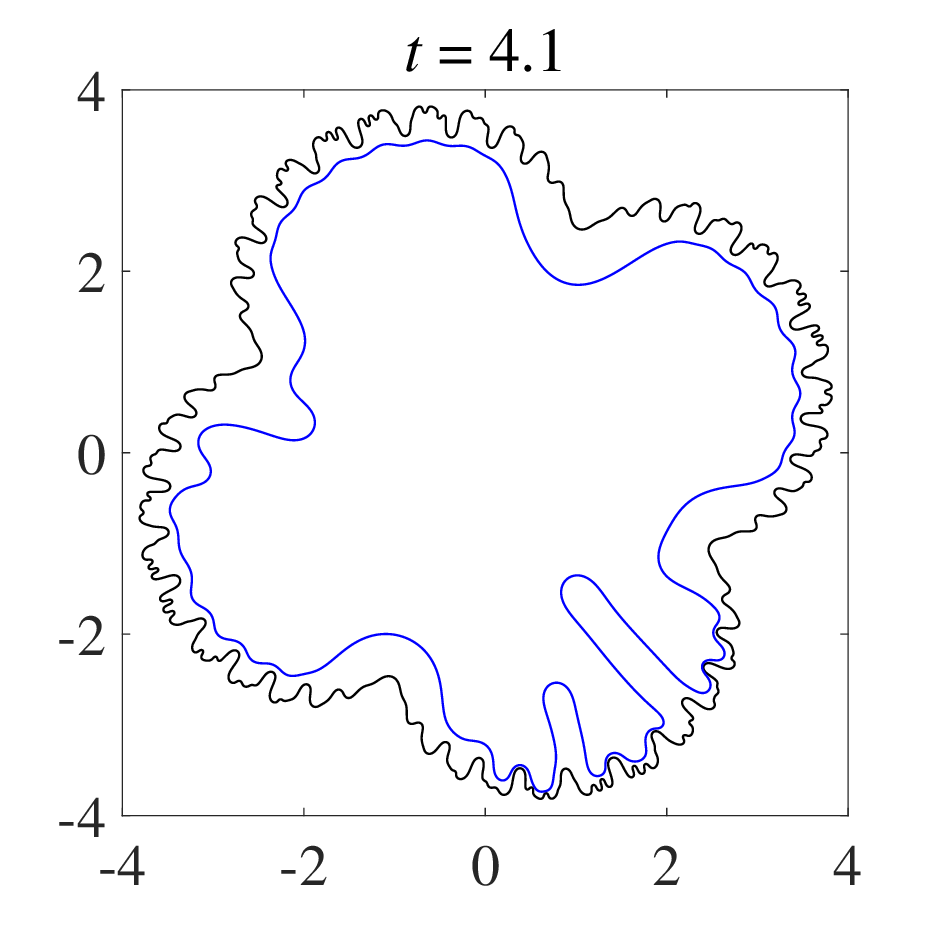}
	
	\caption{Dynamics of the nonlinear interfacial patterns illustrating typical fingering morphologies during three-layer radial Hele-Shaw flows for (a) $R_{0}=0.2$, and (b) $R_{0}=0.5$. The initial conditions considered for the inner and outer interfaces are ${\cal R}_{1}(\theta,0)=1+0.05[\sin(2\theta)+\cos(3\theta)]$ and ${\cal R}_{2}(\theta,0)=1/R_0+ 0.1\sin(4\theta)$, respectively. In a same row, time increases from left to right and all physical parameters are the same as the ones used in Fig.~\ref{int1}.}
	\label{real}
\end{figure}

When studying the behavior of the three-layer flow regarding changes in the parameter $R_{0}$, as we have done in Fig.~\ref{int1}, the initial conditions for the interfaces were composed only by a single cosine mode. In addition, the same mode 4 was used on both the interfaces. Although these considerations are acceptable and allowed us to obtain a number of physical findings though out our study, we would like to close this section by providing a more realistic evolution of the interfaces by taking into account more arbitrary initial conditions. Therefore, following the same layout of presentation used previously in Fig.~\ref{int1}, here in Fig.~\ref{real} we illustrate the temporal evolution of the coupled-interface system for (a) $R_{0}=0.2$ and (b) $R_{0}=0.5$, and utilizing the same set of physical parameters considered in Fig.~\ref{int1} except by the initial conditions: here, the inner interface is a mixture of modes 2 (sine) and 3 (cosine), and the outer interface is composed by a single mode 4 (sine), i.e., ${\cal R}_{1}(\theta,0)=1+0.05[\sin(2\theta)+\cos(3\theta)]$ and ${\cal R}_{2}(\theta,0)=1/R_0+ 0.1\sin(4\theta)$. 

Fig.~\ref{real}(a) illustrates the temporal evolution of the interfaces for $R_{0}=0.2$. The nearly circular interfaces evolve into more complicated patterned structures as time increases. While the outer interface remains almost unchanged with respect to its initial shape, the inner interface changes dramatically, presenting a final convoluted, asymmetric fingering pattern after multiple tip-splitting processes and intense finger competition. For the case with $R_{0}=0.5$ depicted in Fig.~\ref{real}(b), one may note that the formation of a fourfold pattern for both interfaces, presenting small bumps at the tip of each finger. But more importantly, the inner interface is more stable when compared to Fig.~\ref{real}(a) and the nonlinear pattern-forming mechanisms of finger tip-splitting and finger competition are much less intense. 

Finally, it is clear that the same key effects unveiled originally in Fig.~\ref{int1} regarding the role of the fluid annulus on regulating the final morphologies of the interfaces are also present in Fig.~\ref{real}. Moreover, since these two figures are plotted by considering two completely different sets of initial conditions, we can verify explicitly the robustness of our numerical scheme regarding changes in the initial conditions. This strengthens the relevance and generality of our numerical results obtained in our study. As a final remark of our work, we would like to mention that other initial conditions (not reported here) were also utilized and the results are similar to the ones already shown in Figs.~\ref{int1} and~\ref{real}.

\section{Conclusion}
\label{conclude}

In the classical two-fluid radial Hele-Shaw cell problem, a single front spreads radially outward as the interfacial fingers tend to split at their tips and compete among themselves, evolving into a complex branched morphology. In this work, we considered the development of viscous fingering patterns in a three-fluid radial Hele-Shaw cell, where two coupled interfaces are present and the coupling strength is mediated by the distance between them. A recent second-order perturbative analysis of this system~\cite{pedronew} has shown that the initial annulus' thickness has a great impact on the morphologies of the patterns at the weakly nonlinear regime. 

In this paper, we went beyond the weakly nonlinear stage of the dynamics and used a spectrally accurate boundary integral method for simulating the temporal evolution of the dual-interface problem. Our numerical algorithm is second-order accurate in time and allowed us to access the fully nonlinear dynamics of the coupled interfaces in a stable and efficient scheme. By comparing the temporal evolution of the interfacial amplitudes given by our numerical scheme to the evolution as predicted by a second-order mode-coupling theory, we showed the unavoidable necessity of the usage of our fully nonlinear approach in order to get an accurate picture of the morphological elements that arise at the later time regime of the three-layer radial Hele-Shaw flow.

Our numerical results reveal that when the initial annulus' thickness is large, two distinct morphologies for the interfaces are found: a mildly deformed outer interface together with a highly branched inner pattern formed by multiple tip-splitting (and other higher-order ramification) processes and intense finger length variability. However, for a thinner annulus, we found that the interfaces evolve to a similar final shape, which is much less unstable in comparison to the patters for thicker annulus. In particular, we have perceived that finger-tip splitting and finger competition phenomena were evidently restrained on these patterned structures, besides the interesting development of a series of low-amplitude bumps at the tips of the fingers associated with the enhanced growth of high-frequency modes promoted by the increasing coupling strength of interfaces.

It should be pointed out that our numerical findings are qualitatively consistent with similar
types of results obtained in Ref.~\cite{pedronew} through a theoretical perturbative mode-coupling approach. Our numerical patterns are also in good agreement with the experimental findings of Ref.~\cite{cardoso1995formation}. To conclude, it is worthwhile to note that a detailed experimental exploration of such a rich three-fluid radial Hele-Shaw cell system, which considers more general situations like the ones explored here, still needs to be developed in order to shed further light on the impact of initial annulus' thickness on the morphologies of the fingering patters theoretically scrutinized in this work and in Ref.~\cite{pedronew}.

\begin{acknowledgments}
	
S. L. acknowledges the support from the National Science Foundation, Division of Mathematical Sciences grant DMS-1720420. J. L. acknowledges partial support from the NSF through grants DMS-1714973, DMS-1719960, DMS-1763272, and the Simons Foundation (594598QN) for a NSF-Simons Center for Multiscale Cell Fate Research. J. L. also thanks the National Institutes of Health for partial support through grants 1U54CA217378-01A1 for a National Center in Cancer Systems Biology at UC Irvine and P30CA062203 for the Chao Family Comprehensive Cancer Center at UC Irvine.
\end{acknowledgments}

\appendix
\section{Second-order mode-coupling functions}

This appendix presents the expressions for the second-order mode-coupling functions which appear in the text.

In Eq.~(\ref{result1}), the second-order terms are given by

\begin{eqnarray}
F(n,n')= \frac{|n|}{R_{1}} \Bigg \{ \Bigg. \frac{1}{R_{1}^2} \left [ \frac{1}{2} - g_{1}(n,n')~{\rm sgn}(nn') \right ] -\left ( \frac{1}{1 - \beta_{21}} \right)\frac{1}{{\rm Ca} R_{1}^{3}} \left [ 1 - \frac{n'}{2} ( 3 n' + n )  \right ] \Bigg. \Bigg \},
\label{F1}
\end{eqnarray}
\begin{eqnarray}
\label{G1}
G(n,n')=\frac{1}{R_{1}} \left \{ |n| [1 - g_{1}(n,n')~{{\rm sgn}}(nn')] - f_{1}^{-1} \right \},
\end{eqnarray}
\begin{eqnarray}
H(n,n')= \frac{|n|}{R_{2}} \Bigg \{ \Bigg. \frac{1}{R_{2}^2} \left [ \frac{1}{2} - g_{2}(n')~{\rm sgn}(nn') \right ] -\left ( \frac{1}{\beta_{23}-1} \right)\frac{1}{{\rm Ca} R_{2}^{3}} \left [ 1 - \frac{n'}{2} ( 3 n' + n )  \right ] \Bigg. \Bigg \},
\label{H1}
\end{eqnarray}
\begin{eqnarray}
\label{I1}
I(n,n')=\frac{1}{R_{2}} \left \{ |n| [1 - g_{2}(n')~{{\rm sgn}}(nn')] \right \},
\end{eqnarray}
\begin{eqnarray}
\label{J1}
J(n,n')=\frac{|n|}{R_{1}} \Bigg \{ \frac{1}{R_{2}^2}~ \frac{(A_{23}R^{2|n|}+1)R^{(|n'|-|n|)}}{A_{23} (1-R^{2|n'|})} ~{{\rm sgn}}(nn') \Bigg \},
\end{eqnarray}
\begin{eqnarray}
\label{K1}
K(n,n')=\frac{|n|}{R_{1}} \Bigg \{ \frac{(A_{23}R^{2|n|}+1)R^{(|n'|-|n|)}}{A_{23} (1-R^{2|n'|})} ~{{\rm sgn}}(nn') \Bigg \},
\end{eqnarray}
\begin{eqnarray}
\label{L1}
L(n,n')=\frac{|n|}{R_{1}} \Bigg \{ \frac{1}{R_{1}^2}~ \frac{(A_{23}+1)R^{(|n'|+ 2)}}{A_{23} (1-R^{2|n'|})} ~{{\rm sgn}}(nn') \Bigg \},
\end{eqnarray}
\begin{eqnarray}
\label{M1}
M(n,n')=\frac{|n|}{R_{1}} \Bigg \{ \frac{(A_{23}+1)R^{(|n'|+ 2)}}{A_{23} (1-R^{2|n'|})} ~{{\rm sgn}}(nn') \Bigg \},
\end{eqnarray}
where
\begin{eqnarray}
\label{g1}
g_{1}(n,n')= \left( \frac{A_{12}+1}{2A_{12}}\right) \frac{(1+A_{23}R^{2|n|}) (1+R^{2|n'|})}{(1-A_{23}R^{2|n|}) (1-R^{2|n'|})} +\left( \frac{A_{12}-1}{2A_{12}}\right),
\end{eqnarray}
\begin{eqnarray}
\label{g2}
g_{2}(n')=  \frac{A_{23}+1}{A_{23}(1-R^{2|n'|})},
\end{eqnarray}
and the ${\rm sgn}$ function equals $\pm 1$ according to the sign of its argument. 

The second-order expressions in Eq.~(\ref{result2}) are given by

\begin{eqnarray}
{\cal F}(n,n')= \frac{|n|}{R_{1}} \Bigg \{ \Bigg. \frac{1}{R_{1}^2} \left [ \frac{1}{2} - g_{3}(n')~{\rm sgn}(nn') \right ] -\left ( \frac{1}{1 - \beta_{21}} \right)\frac{1}{{\rm Ca} R_{1}^{3}} \left [ 1 - \frac{n'}{2} ( 3 n' + n )  \right ] \Bigg. \Bigg \},
\label{F2}
\end{eqnarray}
\begin{eqnarray}
\label{G2}
{\cal G}(n,n')=\frac{1}{R_{1}} \left \{ |n| [1 - g_{3}(n')~{{\rm sgn}}(nn')] \right \},
\end{eqnarray}
\begin{eqnarray}
{\cal H}(n,n')= \frac{|n|}{R_{2}} \Bigg \{ \Bigg. \frac{1}{R_{2}^2} \left [ \frac{1}{2} - g_{4}(n,n')~{\rm sgn}(nn') \right ] -\left ( \frac{1}{\beta_{23}-1} \right)\frac{1}{{\rm Ca} R_{2}^{3}} \left [ 1 - \frac{n'}{2} ( 3 n' + n )  \right ] \Bigg. \Bigg \},
\label{H2}
\end{eqnarray}
\begin{eqnarray}
\label{I2}
{\cal I}(n,n')=\frac{1}{R_{2}} \left \{ |n| [1 - g_{4}(n,n')~{{\rm sgn}}(nn')] -f_{4}^{-1} \right \},
\end{eqnarray}
\begin{eqnarray}
\label{J2}
{\cal J}(n,n')=\frac{|n|}{R_{2}} \Bigg \{ \frac{1}{R_{2}^2}~ \frac{(A_{12}-1)R^{(|n'|-2)}}{A_{12} (1-R^{2|n'|})} ~{{\rm sgn}}(nn') \Bigg \},
\end{eqnarray}
\begin{eqnarray}
\label{K2}
{\cal K}(n,n')=\frac{|n|}{R_{2}} \Bigg \{ \frac{(A_{12}-1)R^{(|n'|-2)}}{A_{12} (1-R^{2|n'|})} ~{{\rm sgn}}(nn') \Bigg \},
\end{eqnarray}
\begin{eqnarray}
\label{L2}
{\cal L}(n,n')=\frac{|n|}{R_{2}} \Bigg \{ \frac{1}{R_{1}^2}~ \frac{(A_{12}R^{2|n|}-1)R^{(|n'|-|n|)}}{A_{12} (1-R^{2|n'|})} ~{{\rm sgn}}(nn') \Bigg \},
\end{eqnarray}
\begin{eqnarray}
\label{M2}
{\cal M}(n,n')=\frac{|n|}{R_{2}} \Bigg \{ \frac{(A_{12}R^{2|n|}-1)R^{(|n'|-|n|)}}{A_{12} (1-R^{2|n'|})} ~{{\rm sgn}}(nn') \Bigg \},
\end{eqnarray}
where
\begin{eqnarray}
\label{g3}
g_{3}(n')= \frac{A_{12}-1}{A_{12}(1-R^{2|n'|})},
\end{eqnarray}
and
\begin{eqnarray}
\label{g4}
g_{4}(n,n')= \left( \frac{A_{23}-1}{2A_{23}}\right) \frac{(1-A_{12}R^{2|n|}) (1+R^{2|n'|})}{(1+A_{12}R^{2|n|}) (1-R^{2|n'|})} +\left( \frac{A_{23}+1}{2A_{23}}\right).
\end{eqnarray}

\section{Convergence test}
\label{discussion1}

\begin{figure}
	\includegraphics[scale=0.44]{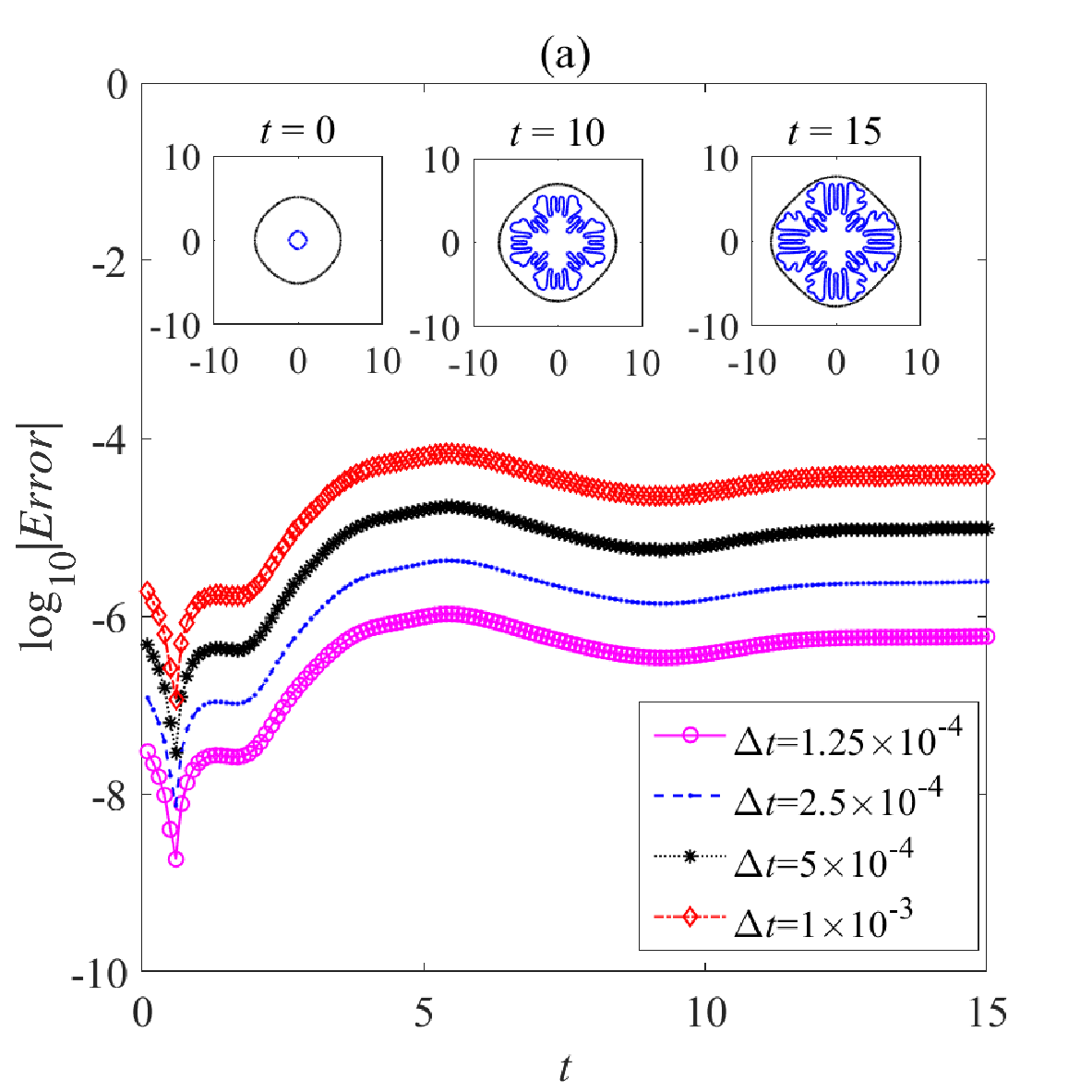}
	\includegraphics[scale=0.44]{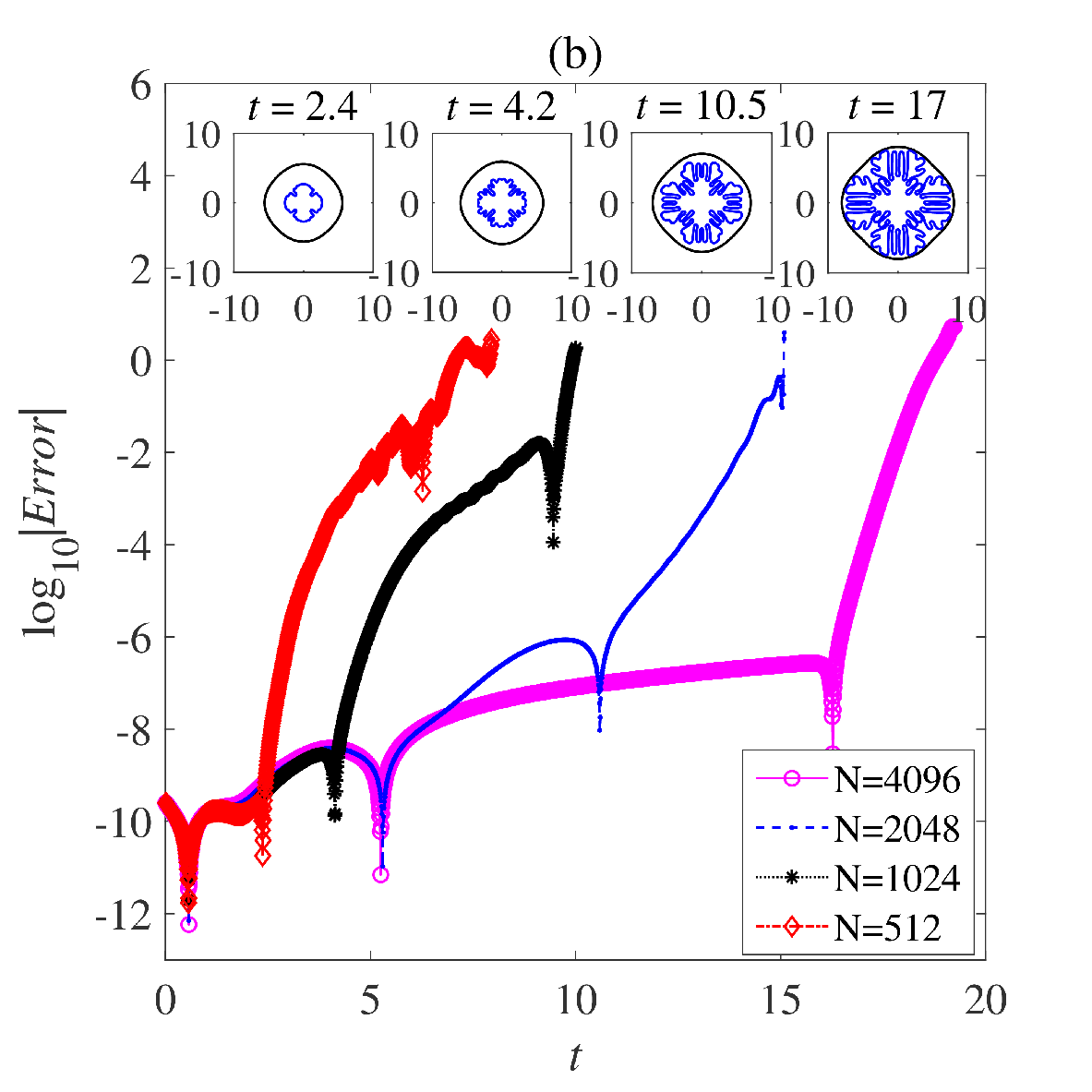}
	\caption{Numerical error $Error=|{A}({t})-{A}(0)|$ plotted as a function of time $t$ for $\beta_{21}=0.01$, $\beta_{23}=100$, ${\rm Ca}=1000$, and $\alpha=1$. In (a) we investigate the temporal convergence by considering $N=4096$ points along each interface and four different values of time step $\Delta t$. In (b) the spatial resolution is analyzed by taking $\Delta t=1\times 10^{-5}$ and utilizing for different values of $N$. In both graphs snapshots of the interfaces are shown as insets.}\label{convergence}
\end{figure}

In this appendix, we show the performance of our numerical scheme, which is $\mathcal{O}(\Delta {t}^2)$ in time and spectral accuracy in space, by simulating the nonlinear dynamics of a three-layer radial Hele-Shaw flow. Therefore, in Fig.~\ref{convergence} the initial interfacial shapes are ${\cal R}_{1}(\theta,0)=1+0.05\cos(4\theta)$ (inner interface) and ${\cal R}_{2}(\theta,0)=5+0.1\cos(4\theta)$ (outer interface), and we take the viscosity ratios as $\beta_{21}=0.01$ and $\beta_{23}=100$, capillary number ${\rm Ca}=1000$, and surface tension ratio $\alpha=1$. According to the weakly nonlinear theory, for these choices of physical parameters both interfaces are unstable and will develop fingering patterns. All computations are performed on a single node in a cluster with 2.8 GHZ CPUs running Linux. 

Due to mass conservation, the area of the annulus domain $\Omega_2$ between the two interfaces should be constant and we use this fact to define the numerical error,  $Error  \equiv |{A}({t})-{A}(0)|$, where $A(t)$ is the area of $\Omega_2$ computed at time $t$ and $A(0)$ is the initial area. To investigate the temporal convergence of our scheme, in Fig.~\ref{convergence}(a) we plot the base 10 logarithm of the temporal error as a function of time for four different values of time step $\Delta t$, and considering $N=4096$ points along each interface. Snapshots of the interfaces are shown as insets for time $t=0$, $t=10$, and $t=15$. As can be seen, lower values of numerical error are associated with smaller values of $\Delta t$. Moreover, the distance between the curves is about 0.6 as the time step is reduced by half, which confirms a second-order accuracy in time. For a fixed time, we have checked that the morphologies depicted by the interfaces are nearly the same regardless of the time step used.

Next perform a spatial resolution study using a fixed time step $\Delta t=1\times 10^{-5}$ and considering different values for the number $N$ of mesh points along each interface. The numerical error is again calculated by the area difference, $Error=|{A}({t})-{A}(0)|$. In Fig. \ref{convergence}(b) we plot the base 10 logarithm of the spatial error versus as a function of time with different resolutions. The curve for $N=512$ points starts differing from the curves using higher resolutions at $t=2.4$. This indicates that more points are needed to resolve the fingering of the inner interface. Similar phenomena also happen at $t=4.2$ for $N=1024$ curve, and at $t=10.5$ for $N=2048$ curve. Using $N=4096$, we have the simulation running up to $t=19.3$. Note that the simulation fails not only because of the highly ramified interface but also because of the close distance between the two interfaces. The smallest distance between interfaces $d$ is about $0.089$ while the grid size of the inner interface is about $0.06$. This might lead to a nearly singularity of the integrals in Eqs.~(\ref{eqmu1}) and (\ref{eqmu2}). Four sample morphologies of the interfaces are shown as insets. We notice that all these simulations produce almost identical numerical results at the same time, indicating spectral accuracy in space. 

\end{document}